\documentclass[journal]{IEEEtran}
\usepackage[utf8]{inputenc}

\usepackage{xcolor}
\usepackage{color, soul}
\usepackage{enumerate}
\usepackage{hyperref}
\usepackage{cite}
\usepackage{tabu}
\usepackage{bm, bbm}
\usepackage{mathtools}
\usepackage{amsmath, amssymb, amsthm}
\usepackage{nicefrac}
\usepackage{empheq}

\usepackage{booktabs}
\usepackage{graphicx, caption}
\usepackage{multirow}
\usepackage{units}
\usepackage{siunitx}
\usepackage{todonotes}
\usepackage{enumitem}

\usepackage{placeins}

\newcommand{\subparagraph}{}
\usepackage{titlesec}

\theoremstyle{definition} 
\newtheorem{prop}{Proposition}
\theoremstyle{definition} 

\theoremstyle{remark} 
\newtheorem{remark}{Remark}

\newenvironment{ldescription}[1]
  {\begin{list}{}%
   {\renewcommand\makelabel[1]{##1\hfill}%
   \settowidth\labelwidth{\makelabel{#1}}%
   \setlength\leftmargin{\labelwidth}
   \addtolength\leftmargin{\labelsep}}}
  {\end{list}}

\newcommand{\set}[1]{\mathcal{#1}} 
\newcommand{\eps}{\bm{\epsilon}}

\newcommand{\myproofstart}{\noindent\textit{Proof. }}
\newcommand{\myproofend}{\hspace*{\fill} $\square$ \vspace{0.2\baselineskip}\\}

\DeclareMathOperator{\Var}{Var}

\DeclareMathOperator{\Tr}{Tr}
\DeclareMathOperator{\diag}{diag}

\DeclareMathOperator{\Eptn}{\mathbb{E}}

\DeclarePairedDelimiter{\scal}{\langle}{\rangle}

\usepackage{amsmath,amsfonts,amsthm}

\title{Online Learning for Network Constrained Demand Response Pricing in Distribution Systems}
\author{Robert Mieth, \textit{Student Member, IEEE} and Yury Dvorkin, \textit{Member, IEEE}. 
}
\begin{document}
\pagestyle{empty}
\bstctlcite{IEEE:BSTcontrol} 
\maketitle

\begin{abstract}
Flexible demand response (DR) resources can be leveraged to accommodate the stochasticity of some distributed energy resources. 
This paper develops an online learning approach that continuously estimates price sensitivities of residential DR participants and produces such price signals to the DR participants that ensure a desired level of DR capacity. 
The proposed learning approach incorporates the dispatch decisions on DR resources into the distributionally robust chance-constrained optimal power flow (OPF) framework.
This integration is shown to adequately remunerate DR resources and  co-optimize the dispatch of DR and conventional generation resources.
The distributionally robust  chance-constrained formulation only relies on  empirical data acquired over time and makes no restrictive assumptions on the underlying distribution of the demand uncertainty. The distributional robustness also  allows for robustifying the otpimal solution against systematically misestimating  empirically learned parameters. 
The effectiveness of the proposed learning approach is shown via numerical experiments. The paper is accompanied by the code and data supplement released for public use.
\end{abstract}


\section*{Nomenclature}

\label{sec:nomenclature}

\noindent
\emph{Sets:}
\vspace{1mm}
\begin{ldescription}{$xxxx$}
\item [$\set{N}$] Set of nodes, indexed by $i =\{0,1,\ldots, n \}$, \\ $|\set{N}|= n+1 =\vcentcolon m$
\item [$\set{N}^+$] Set of nodes without the root node, i.e. $\set{N}^+ =\{\set{N} \backslash 0 \}$
\item [$\set{L}$] Set of edges/lines indexed by $i\in\set{N}^+$
\item [$\set{G}$] Set of controllable generators, \mbox{$\set{G}\subseteq\set{N}$}
\item [$\set{A}_i$] Set of ancestor nodes of node $i$
\item [$\set{C}_i$] Set of children nodes of node $i$
\item [$\set{T}$] Set of time intervals, indexed by $t$
\item [$\Lambda_t$] Set of historic price signals at time $t$
\item [$\set{X}_t$] Set of historic demand response observations at time $t$
\end{ldescription}

\noindent
\emph{Variables and Parameters:}
\vspace{1mm}
\begin{ldescription}{$xxxx$}
\item [$c_i(\cdot)$] Cost function of the generator at node $i$
\item [$\bar{d}_{i,t}^{P}$] Active power demand forecast at node $i$ at time~$t$
\item [$\bar{d}_{i,t}^{Q}$] Reactive power demand forecast at node $i$ at time~$t$
\item [$e$] Column vector of ones of appropriate dimensions
\item [$f^{P}_{i,t}$] Active power flow on line $i$ towards node $i$ at time~$t$
\item [$f^{Q}_{i,t}$] Reactive power flow on line $i$ towards node $i$ at time~$t$
\item [$g_{i,t}^{P}$] Active power  output  at node $i$ at time~$t$
\item [$g_{i,t}^{Q}$] Reactive power  output  at node $i$ at time~$t$
\item [$g_{i,t}^{P,\min},g_{i,t}^{P,\max}$] Minimum/Maximum active power output
\item [$g_{i,t}^{Q,\min},g_{i,t}^{Q,\max}$] Minimum/Maximum reactive power output
\item [$s$] Auxiliary variable for distributionally robust chance-constraints reformulation
\item [$u_{i,t}$] Voltage squared at node $i$ at time $t$
\item [$w_i(x)$] Cost/discomfort of demand reduction $x$
\item [$x_{i,t}$] Active power demand reduction at node $i$ at time~$t$
\item [$x^*_{i,t}$] Optimal/desired $x_{i,t}$
\item [$A$] Mapping of net-injections to line flows, $\mathbb{R}^{n\times m}$
\item [$B_t$] Parameter estimation Error at time $t$
\item [$C$] Auxiliary matrix $(\alpha e^\top - I)\in\mathbb{R}^{m\times m}$ that maps $\epsilon$ into changes in nodal injections
\item [$F_{i,t}$] Fisher information of node $i$ at time $t$
\item [$I$] Identity matrix of appropriate dimensions
\item [$L_{i,t}$] Sum of squared price differences at $i$ from mean at $t$
\item [$M$] Auxiliary matrix $\mathbb{R}^{m+1\times m+1}$ of decision variables 
\item [$R_i$] Resistance of line $i$
\item [$R$] Auxiliary diagonal matrix $\diag[R_i, \forall  i \in \set{N}^+] \in \mathbb{R}^{n\times n}$
\item [$S_i$] Apparent power on line $i$
\item [$T_i(\alpha)$] Mapping of load error vector $\epsilon_t$ to voltage change at node $i$ as a function of  vector $\alpha$ 
\item [$X_i$] Reactance of line $i$
\item [$X$]  Auxiliary diagonal matrix $\diag[X_i, \forall  i \in \set{N}^+] \in \mathbb{R}^{n\times n}$
\item [$\alpha_{i,t}$] Participation factor of the generator at node $i$ at time~$t$
\item [$\alpha_t$] Auxiliary vector $\{\alpha_{i,t}, i \in \set{N}\} \in \mathbb{R}^{m\times1}$ 
\item [$\beta_i$] Parameters of the price sensitivity model  at node $i$, \mbox{$\beta_i \coloneqq \{\beta_{0,i}, \beta_{1,i}\}$}
\item [$\hat{\beta}_{i}^{(t)}$] Estimation of  $\beta_i$ at time $t$, \mbox{$\hat{\beta}_i^{(t)} \coloneqq \{ \hat{\beta}_{0,i}^{(t)}, \hat{\beta}_{1,i}^{(t)}\}$}
\item [$\gamma_{i,t}$] Ratio between the active and reactive power demand
\item [$\gamma_t$] Auxiliary diagonal matrix $\diag[\gamma_{i,t}, i \in \set{N}^+] \!\in\! \mathbb{R}^{n\times n}$
\item [$\eps$] Random vector of disturbance of the expected reduced load \mbox{$\eps \coloneqq \{ \eps_{i}, i \in \set{N} \} \in \mathbb{R}^{m\times1}$}
\item [$\epsilon_t$] Realization of $\eps$ at time $t$, $\epsilon_t \coloneqq \{\epsilon_{i,t}, i \in \set{N}\} \in \mathbb{R}^{m\times1}$
\item [$\hat{\epsilon}_{\tau}^{(t)}$] Vector of residual errors at time interval $\tau$ based on the parameter estimation in time $t$, ${\hat{\epsilon}_{\tau}^{(t)} \coloneqq \{\hat{\epsilon}_{i,\tau}^{(t)}, i \in \set{N}\} \in \mathbb{R}^{m\times1}}$
\item [$\eta$] Acceptable likelihood of constraint violations
\item [$\kappa_t$] Fixed retail tariff per unit of active power  at time $t$
\item [$\lambda_{i,t}$] Price signal for node $i$ at time~$t$
\item [$\hat{\mu}_t$] Empirical mean of residuals at time $t$
\item [$\nu_{0,i},\nu_{1,i}$] DR participation cost parameters 
\item [$\sigma^2_i$] Load variance at node $i$
\item [$\omega_t$] Energy price at substation at time $t$
\item[$\zeta(t)$] Regret at time $t$
\item [$\Pi_i(x)$] Profit of participating providing demand reduction $x$
\item [$\hat{\Sigma}^{(t)}$] Empirical covariance matrix of residuals at time $t$
\item [$\hat{\Omega}^{(t)}$] Empirical second order moment matrix of residuals at time $t$
\end{ldescription}
\noindent
The \textbf{bold} characters indicate random variables/parameters.

\noindent
\emph{Operators:}
\vspace{1mm}
\begin{ldescription}{$xxxxx$}
\item [${\mathbb{E}[\cdot]}$] Expectation of a random variable
\item [${\mathbb{P}[\cdot]}$] Probability of an event
\item [$X^\top$] Transpose of matrix $X$ 
\item [$\scal{X,Y}$] Trace product of matrices, i.e. $\scal{X,Y}= \Tr(XY)$
\item [$\Theta(\cdot)$] ``Big-Theta'' notation for complexity of algortihms
\end{ldescription}


\section{Introduction}
\label{sec:introduction}

Leveraging flexible distributed loads via  demand response (DR) programs allows electric power distribution utilities to mitigate the volatility of intermittent distributed energy resources (DERs), reducing peak loads, and avoiding electricity surcharges for customers \cite{qdr2006benefits}. Such programs mainly target commercial and industrial loads that are relatively homogeneous in size and technical capabilities and, thus, are fairly easy to price and interface with energy managements systems used by utilities \cite{siano2014demand}. On the other hand, enrolling residential-scale DR resources is challenging due to their heterogeneous characteristics and electricity usage patterns and preferences. 
For example, Consolidated Edison of New York  has recently introduced its voluntary ``Smart Air Conditioner'' program \cite{coned_smartac}. During peak demand hours, the app-based system requests permission to adjust temperature setting of residential air conditioning units via a WiFi-connected module. In return, residents receive a certain amount of ``points'', which can be redeemed as retail gift cards. However, this program does not differentiate the DR participants and, therefore, cannot provide customized incentives to accurately match participant preferences and utility needs. This paper develops an online learning approach that estimates price sensitivities of residential DR participants and produces the price  signals that ensures a desired DR capacity. 

{\color{black} 
Existing incentive-based DR programs, e.g. \cite{deng2015survey,sarker2015optimal,conejo2010real_formatted}, optimize the amount of demand reduction needed by the system and price signals in a look-ahead manner. 
However, these approaches do not guarantee that the observed response of DR participants meets the expectation because there is no feedback communication channel from the DR participants to the utility.
However, explicitly surveying price sensitivities or two-way \textit{a priori} negotiation incurs a large communication overhead and may expose sensitive data such as consumption habits. Alternatively, utilities  may prefer one-way (passive) approaches to learn consumption patterns and preferences of individual DR participants indirectly \cite{gomez2012learning}. Such indirectly collected data can suffer from various inaccuracies, thus also introducing uncertainty on the deliverable DR capacity. 
}

To realistically estimate the response of each DR participant and reduce its uncertainty,   \cite{Li_A_2017,Khezeli_Risk_2017,gomez2012learning,jia2013day} develop online learning methods based on continuous regression. These methods learn the price sensitivity of each DR participants by inferring it from the historical price signals and observed DR responses, and use the inferred value to generate a more accurate price signal. 
Li \textit{et al.} \cite{Li_A_2017}  use an iterative regression algorithm to learn price sensitivities of individual DR participants that can be used by profit-seeking DR aggregators to optimize the total DR capacity offered to the utility. This algorithm is shown to achieve logarithmically progressing regret, i.e. the deviation from the perfect foresight case as a function of the learning horizon. 
Similarly to \cite{Li_A_2017}, Khezeli \textit{et al.} \cite{Khezeli_Risk_2017} develop a risk-averse learning approach for utilities operating residential DR programs, which can provide an explicit probabilistic guarantee on the anticipated payoff of utilities.
Jia \textit{et al.} \cite{jia2013day} develop a learning algorithm that allows for a utility or an DR aggregator to participate in a two-stage (day-ahead and real-time) whole-sale market. The proposed learning algorithm also has logarithmic regret over the learning horizon and is used to obtain the aggregated demand function of the DR participants to optimize the wholesale bidding strategy and arbitrage between the day-ahead and real-time stages. The common limitation of  \cite{Li_A_2017,Khezeli_Risk_2017,gomez2012learning,jia2013day} is that distribution network constraints, e.g. nodal voltage and line flow limits, are ignored, which can reduce deliverability of  DR capacity in practice.

Modeling network constraints for distribution systems requires considering AC power flows to accurately account for both voltage magnitudes and line flows. Since AC power flow equations are NP-hard, \cite{7063278}, one can use relaxation \cite{6756976} or linearization \cite{doi:10.1287/ijoc.2014.0594} techniques for the sake of computational tractability.  Additionally, the effect of uncertain nodal injections on voltage magnitudes and line flows must be accounted for. To avoid dealing with computationally demanding scenario-based stochastic programming, \cite{DallAnese_Chance_2017, dr-ac-ccopf, hassan2018optimal} use the chance constrained framework. 
Dall'Anese \textit{et al.} \cite{DallAnese_Chance_2017} and Hassan \textit{et al.} \cite{hassan2018optimal} formulate  an AC Chance-Constrained Optimal Power Flow (CC-OPF) problem by linearizing AC power flow equations and using given assumptions on the underlying uncertainty sources.
Mieth \textit{et al.} \cite{dr-ac-ccopf} extend the formulation in  \cite{DallAnese_Chance_2017} to a distributionally robust CC-OPF (DRCC-OPF) formulation that immunizes its solution against a family of uncertainty distributions drawn from empirical data. 

\begin{figure}
    \centering
    \includegraphics[width=0.9\linewidth]{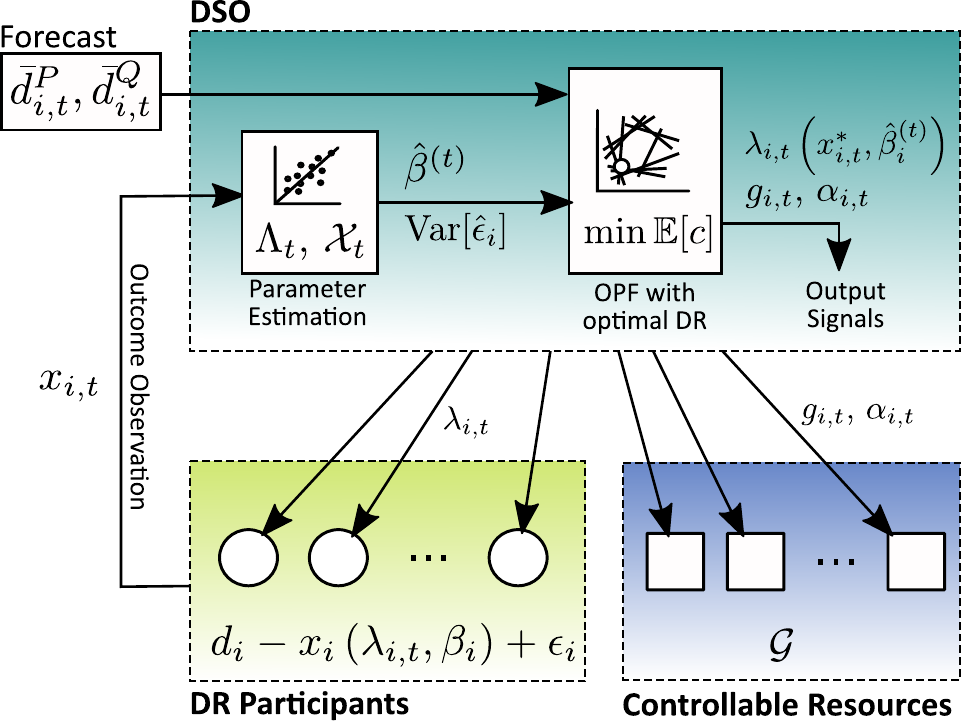}
    \caption{The proposed online learning approach in a distribution system with DR participants and controllable resources.}
    \label{fig:dr_system}
\end{figure}

{\color{black} 
This paper aims to bridge the gap between online learning methods for estimating the price sensitivities of DR participants and the DRCC-OPF framework. 
The developed online learning method is a dynamic pricing scheme \cite{den2013simultaneously, keskin2014dynamic} that optimizes price signals for DR participants with unknown prices and and co-dispatches the DR and system resources as shown in Fig.~\ref{fig:dr_system}.
}
Given the price signals, the utility observes the response of  DR participants and updates its knowledge of price sensitivities. Relative to \cite{Li_A_2017,Khezeli_Risk_2017,gomez2012learning,jia2013day}, this paper internalizes the effects of network constraints and distributionally robust optimization on learning.
Distributional robust optimization mitigates risk imposed by incomplete information on DR parameters and underlying uncertain disturbances. By explicitly treating risk as part of the optimization, the model will  learn both the DR price sensitivities and the distribution of the load disturbances.
Furthermore, by relying on the empirical distribution this work generalizes the approach of \cite{dr-ac-ccopf} towards uncertain load errors that are potentially non-Gaussian and correlated. 

\section{DR Model for Learning Price Sensitivities}
\label{sec:problem_formulation}

This section describes the proposed DR model from the perspective of the utility. We adopt the current US practice, where a single Distribution System Operator (DSO) controls the entire distribution system and possesses all measurements. Specifically, it is assumed that the DSO can characterize every time interval $t \in \set{T}$ with set  $\Lambda_t =\{\lambda_\tau \in \mathbb{R}^m, \forall \tau \leq t-1\}$, where $\lambda_\tau$ is the vector of price signals sent to the DR participants at preceding time intervals, and with set $\set{X}_t =\{x_\tau \in \mathbb{R}^{m}, \forall \tau \leq t-1\}$, where $x_\tau$ is the vector of observed DR responses. For simplicity, it is assumed that every node of the distribution system hosts one participant that represents the aggregated behavior of all participants connected to that node and, therefore, vectors $\lambda_\tau$ and $x_\tau$ can further be itemized for every node such that $\lambda_t =\{\lambda_{i,t} \in \mathbb{R}, \forall i \in \set{N}\}$ and $x_t =\{x_{i,t} \in \mathbb{R}, \forall i \in \set{N}\}$, respectively. Additionally, the utility possess  nodal active and reactive net demand forecasts $\bar{d}_t^P = \{\bar{d}_{i,t}^P \in \mathbb{R}, \forall i \in \set{N}\}$ and $\bar{d}_t^Q = \{\bar{d}_{i,t}^Q \in \mathbb{R}, \forall i \in \set{N}\}$.

Using this information, the DSO acts as follows: 
\begin{enumerate}[topsep=2pt]
    \item It aims to maximize the expected operating cost considering the cost of electricity provision, remuneration for DR participants and revenues from selling energy to consumers.
    \item It determines the  dispatch of all dispatchable DERs (e.g. power outputs of controllable resources and the amount of reserve they provide) and ensures that all distribution system constraints are met.
    \item It generates the DR price signal to achieve a  desirable response from the DR participants.  
\end{enumerate}

\subsection{Price Sensitivity Model}
\label{ssec:price_senstivity_model}
At time $t$ the DR participant at node $i$ receives the price signal $\lambda_{i,t}$ and has to decide on the amount of demand reduction $x_{i,t}$ that satisfies the trade-off between receiving the remuneration $\lambda_{i,t} x_{i,t}$ and the lost utility of not consuming $x_{i,t}$. 
{\color{black} 
Assume that the cost (or lost utility) $w_i(x_{i,t})$ of providing demand response $x_{i,t}$ follows a quadratic function \cite{deng2015survey} so that  
\begin{align}
    w_i(x_{i,t}) = \frac{1}{2} \nu_{1,i} x_{i,t}^2 + \nu_{0,i} x_{i,t}, 
\end{align}
where $\nu_{1,i}$, $\nu_{0,i}$ are participant-specific parameters.
The profit maximization problem at each node is therefore given by:
\begin{align}
    \max_{x_{i,t}} \Pi_i(x_{i,t}) \coloneqq  \lambda_{i,t} x_{i,t} - w_i(x_{i,t}).
\end{align}
Under first order optimality conditions, $\Pi_i$ is maximized if 
\begin{align}
    x_{i,t}^* = \frac{1}{\nu_{1,i}} \lambda_{i,t} - \frac{\nu_{0,i}}{\nu_{1,i}},
\end{align}
which motivates the choice of  the following  linear DR model:
\begin{align}
    x_{i,t}(\lambda_{i,t}) = 2 \beta_{1,i} \lambda_{i,t} + \beta_{0,i}, \label{eq:result}
\end{align}
where $\beta_{1,i} = \frac{1}{2\nu_{1,i}}$ and $\beta_{0,i}  =  -\frac{\nu_{0,i}}{\nu_{1,i}}$. Similarly to our result in \eqref{eq:result},  \cite{besbes2015surprising} shows that   linear models  fit  the majority of price-sensitive demand models, because their dispatchable range  is typically small and can be approximated linearly,  \cite{Khezeli_Risk_2017,huang2019transactive}. One can additionally limit the available DR amount by enforcing an upper bound to restrict the dispatchable DR range to its linear segment, e.g. similarly to \eqref{eq:dr_limit} below. 
}

Due to various externalities (e.g. some short-term adaptations of comfort-level constraints \cite{7961263}), the reaction of  DR participants to price signal $\lambda_{i,t}$ will be subject to random deviations (\textit{noise}). As a result, the demand reduction observed by the utility can be represented by random variable  $\bm{x}_i(\lambda)$, which relates the price signal and uncertain DR capacity, with the following expectation and variance:
\begin{align}
    \mathbb{E}[\bm{x}_i(\lambda)] &= h(\beta_i, \lambda) = 2\beta_{1,i}\lambda + \beta_{0,i}  \label{eq:model_expected_dr}
    \end{align}
    \begin{align}
    \Var[\bm{x}_i(\lambda)] &= \sigma_i^2\textbf{}
    \label{eq:model_dr_var}
\end{align}
\noindent
where $\beta_i = \{\beta_{0,i}, \beta_{1,i}\}, \forall i \in \set{N},$ are unknown parameters that the DSO needs to learn. 
In terms of the physical interpretation of this model, parameter $\beta_{0,i} = 0, \forall i \in \set{N},$ ensures that there is no DR for $\lambda_{i,t}=0$ and  parameter $\beta_{1,i} \geq 0, \forall i \in \set{N},$ so that the  amount of demand reduction is (weakly) increasing as $\lambda_{i,t}$ increases. 
The variance of the observed DR capacity in~\eqref{eq:model_dr_var} is  constant within a given price range, since it depends on characteristics of the DR participant and does not typically exhibit any noticeable sensitivity to the price signal \cite{besbes2015surprising}.  

Given \eqref{eq:model_expected_dr}, the  error of the observed demand reduction is:
\begin{align}
    \eps_{i} \coloneqq \bm{x}_{i}(\lambda) - \mathbb{E}[\bm{x}_{i}(\lambda)].
    \label{eq:error_definition}
\end{align}
\noindent
As $\mathbb{E}[\bm{x}_{i}(\lambda)]$ is the expected value of $\bm{x}_{i}(\lambda)$,  \mbox{$\mathbb{E}[\eps_{i}] = 0$} and $\Var[\eps_{i}] = \Var[\bm{x}_i(\lambda)]=\sigma_i^2$ by definition. 

{\color{black} 
\begin{remark}
While parameter $\beta_{0,i}$ in \eqref{eq:result}  is set to zero due to the physical interpretation of the price signal, i.e.  $\lambda_{i,t} = 0$  when  $x_{i,t}=0$, we model $\beta_{0,i} \neq 0$ to provide an additional degree of freedom for the parameter estimation process. If $\beta_{0,i} \neq 0$, it captures systematic errors due to the imperfection of demand forecasting and learning.
\end{remark}
}

\subsection{Observable Demand Response Error}
\label{ssec:closer_look_on_error_term}

As per the model in \eqref{eq:model_expected_dr}, the total demand observed by the DSO at time $t$ is given as:
\begin{align}
    d_{i,t}^P = \bar{d}_{i,t}^P - h(\lambda_{i,t}, \beta_i) - \epsilon_{i,t}
    \label{eq:resulting_demand}
\end{align}
where $\bar{d}^P_{i,t}$ is the forecast  demand at node $i$ at time $t$, $h(\lambda_{i,t}, \beta_i)$ is the true DR expectation based on unknown parameter $\beta_i$ and price signal $\lambda_{i,t}$, and error $\epsilon_{i,t}$ is a given realization drawn from random vector $\eps_i$.
Since the DSO can only observe  the total difference between the forecast and actual demand, $\epsilon_{i,t}$ includes both the forecast error of $\bar{d}_{i,t}^P $ and the DR noise of $h(\lambda_{i,t}, \beta_i)$. Using this aggregated error, we recover the observed DR capacity as:
\begin{align}
    x_{i,t} = \bar{d}_{i,t}^P - d_{i,t}^P, \label{eq:oneequation_epsilon}
\end{align} 
where  $x_{i,t}$ internalizes aggregated demand variance regardless of its cause, which  can be included in the DRCC-OPF  below.
{\color{black} 
In practice, net demand observations $d_{i,t}^P$ for each time step~$t$ will be obtained from SCADA or user-end smart meter measurements. 
Since a typical temporal resolution of these measurements (subseconds to minutes) is smaller than the resolution of DR programs (minutes to hours) \cite{dr_advance_metering}, random measurement errors can be mitigated by simple filtering, e.g. averaging, \cite{oppenheim2015signals}. 
Since such a filtering procedure can be executed as a pre-processing step for the proposed learning scheme, $d_{i,t}^P$ represents refined measurements. 
}

Observing true disturbance $\epsilon_{i,t}$, however, is impossible without knowing the true expectation of $\bm{x}_i$, i.e. knowledge of  true parameters $\beta_i$. Therefore, the DSO only observes the \textit{residual} error that can be computed as follows:
\begin{align}
\hat{\epsilon}_{i,\tau}^{(t)} = x_{i,\tau} - h(\lambda, \hat{\beta}_i^{(t)}) \qquad \forall \tau \leq t-1,
\label{eq:ref_residual_def}
\end{align}
where $\hat{\beta}_i^{(t)}$ is the estimate of the price sensitivity parameters of node $i$ available to the DSO at time $t$. If the estimate was perfect, i.e. $\hat{\beta}_i^{(t)} = \beta_i$, the residual error $\hat{\epsilon}_{\tau,i}^{(t)}$ would be equal to  true disturbance $\epsilon_{\tau,i}$ for all previous timesteps ${\tau\in\{1,\ldots, t-1\}}$. 

\subsection{Learning Price Sensitivities} 
\label{sec:parameter_estimation} 
{\color{black} 
At each time step the DSO computes estimates $\hat{\beta}^{(t)}_{0,i}$  and $\hat{\beta}^{(t)}_{1,i}$ of unknown parameters $\beta_{0,i}$, $\beta_{1,i}$ to update price sensitivity model $h(\lambda, \hat{\beta}^{t}_i)$ and to evaluate the residual error in \eqref{eq:ref_residual_def}.
These estimates can be obtained from  historical observations $\Lambda_t$ and $\set{X}_t$ using the least-square estimator (LSE) as follows: 
}
\begin{align}
    \hat{\beta}^{(t)}_{1,i} &= \frac{\sum_{\tau=1}^{t-1}(\lambda_{i,\tau} - \bar{\lambda}_{i,t})(x_{i,\tau} - \bar{x}_{i,t})}{2\sum_{\tau=1}^{t-1}(\lambda_{i,\tau} - \bar{\lambda}_{i,t})^2} 
    \label{eq:lse_beta_1}
    \end{align}
  \begin{align}  
    \hat{\beta}^{(t)}_{0,i} &= \bar{x}_{i,t} - \hat{\beta}_{1,i}\bar{\lambda}_{i,t},
    \label{eq:lse_beta_2}
\end{align}
\noindent
where  \eqref{eq:lse_beta_1} and \eqref{eq:lse_beta_2} are derived in Appendix~\ref{ax:lse_derivation} and:
\begin{align}
    \bar{\lambda}_{i,t} = \frac{1}{t-1} \sum_{\tau=1}^{t-1} \lambda_{i,\tau}, \quad
    \bar{x}_{i,t} = \frac{1}{t-1} \sum_{\tau=1}^{t-1} x_{i,\tau}.
\end{align}
{\color{black} 
The LSE approach fundamentally matches the price sensitivity model \eqref{eq:result} and using $\hat{\beta}^{(t)}_{1,i}$ and $\hat{\beta}^{(t)}_{0,i}$, we obtain the expected DR participation $h(\lambda, \hat{\beta}_i^{(t)})$ as a function of $\lambda$ based on the available historical data.
After estimating $h(\lambda, \hat{\beta}_i^{(t)})$, we obtain set of residual vectors \mbox{$\hat{\set{E}}_t = \{\hat{\epsilon}_1^{(t)}, \hat{\epsilon}_2^{(t)},\ldots, \hat{\epsilon}_{t-1}^{(t)}\}$}, where each element $\hat{\epsilon}^{(t)}_\tau$ is a vector of nodal residual errors from \eqref{eq:ref_residual_def} for each time $t$, i.e.  $\hat{\epsilon}^{(t)}_\tau = \{\hat{\epsilon}_{i,\tau}^{(t)}, \forall i \in \set{N}\}$.
} 
As the learning procedure progresses, set $\hat{\set{E}}_t$ is updated at every time~$t$ because its elements depend on the value of parameters $\hat{\beta}^{(t)}_{0,i}$ and $\hat{\beta}^{(t)}_{1,i}$ obtained at time $t$.  

The residual errors, estimated as described above, are then used to characterize random vector $\eps$ in an empirical manner, i.e. based on the observations collected by the DSO. This residual-error-centric approach has multiple advantageous properties. 
{\color{black} 
First, since the random error is independent from the price signal, the LSE method yields that the expected value of the residual error observed by the DSO is zero, i.e. 
    $\Eptn[\hat{\epsilon}_{t} | \Lambda_t, \set{X}_t] = 0.$}
Note that this property is obtained by not restricting $\hat{\beta}_{0,i}$ to zero but allowing the estimator to find the minimal error with all possible degrees of freedom.
Second, at every time interval~$t$ the empirical mean vector $\hat{\mu}^{(t)}$ and empirical covariance matrix $\hat{\Sigma}^{(t)}$ can be computed as:
\allowdisplaybreaks
\begin{align}
    &\hat{\mu}^{(t)} = \frac{1}{t-1} \sum_{\tau=1}^{t-1} \epsilon_{\tau}  \label{eq:emp_mean}\\
    &\hat{\Sigma}^{(t)} \!=\! \frac{1}{t-2} \!\sum_{\tau=1}^{t-1} (\hat{\epsilon}_{\tau}\! -\! \hat{\mu}^{(t)})(\hat{\epsilon}_{\tau} \!-\! \hat{\mu}^{(t)})
     \quad \forall i,j \in \set{N}.  \label{eq:emp_var} 
\end{align}%
\allowdisplaybreaks[0]%
These  parameters $\hat{\mu}^{(t)}$  and $ \hat{\Sigma}^{(t)}$ can be leveraged toward the DRCC-OPF described below. 
Using these characteristics of the empirical distribution allows to overcome the limitation of making specific assumptions on the true underlying distribution (e.g. Gaussian as in \cite{Bienstock_Chance_2014, DallAnese_Chance_2017, dr-ac-ccopf,hassan2018optimal}). Rather, learning can be performed over empirical data sets, while fully accounting for spatio-temporal sensitivities captured in the covariance matrix. In the context of DR participant scattered across a given distribution system, such sensitivities are particularly self-manifesting due to similar external conditions. 
The LSE above can be adapted to deal with time-variable behavior of DR participants.  
{\color{black} 
For instance, if the price sensitivity varies across a given day (e.g. morning, afternoon, night), different sets of sensitivities can be learned for different time periods.
This way, imperfections of linear response functions (e.g. minimum and maximum cut-off DR regions)  can be mitigated. Furthermore, allowing for $\hat{\beta}_{0,i} \neq 0$ also contributes to improving the estimation of potential nonlinearites close to the bounds of the response domain (e.g. close to saturation), \cite{chatterjee2013handbook}. 
}
Further, the LSE can be adapted to either discard older data points or to weight relatively recent data points higher than older  ones to capture systematic sensitivity changes. We consider time-invariant price sensitivities below.

\section{Network Model}

\begin{figure}[!b]
\centering
\includegraphics[width=0.7\linewidth]{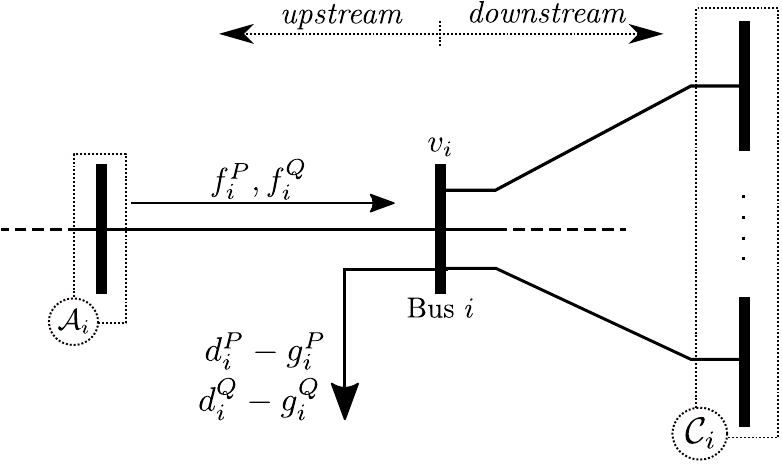}
\caption{\small Power flow notations in a radial distribution network.}
\label{fig:notation_illustration}
\end{figure}

We consider a radial distribution system, as shown in {\color{black} Fig.}~\ref{fig:notation_illustration}, 
 represented by graph $\Gamma (\set{N}, \set{L})$, where $\set{N}$ and $\set{L}$ are the sets of nodes indexed by $i \in \{0,1,\ldots,n\}$ and edges (lines) indexed by $i\in\{1,2,\ldots,n\}$, respectively. The graph is a tree with the root node indexed as $0$ and $\set{N}^+ \coloneqq \set{N} \setminus \{0\}$ is the set of all non-root nodes. All nodes have one ancestor  node $\set{A}_i$ and a set of children nodes $\set{C}_i$. Since $\Gamma $ is radial, it is \mbox{$|\set{A}_i| = 1, \forall i \in \set{N}^+$} and all edges $i \in \set{E}$ are indexed by $\set{N}^+$. The root node (substation) connects the distribution system and the transmission system, where electricity is supplied at wholesale price $\omega_t$. Each node is characterized by its net demand injection of active and reactive power ($d_{i,t}^P$ and $d_{i,t}^Q$, $i \in \set{N}$), and its voltage magnitude $v_{i,t} \in [v_{i,t}^{\min}, v_{i,t}^{max}], i \in \set{N}$, where $v_{i,t}^{\min}$ and $v_{i,t}^{\max}$ are the upper and lower  limits.  To use linear operators, we introduce $u_{i,t} = v_{i,t}^2, i \in \set{N}$. If a node is equipped with a controllable DER (e.g. cogeneration resource), we model active and reactive power generation $g_{i,t}^P \in [g_{i,t}^{P,\min}, g_{i,t}^{P,\max}]$ and $g_{i,t}^Q \in [g_{i,t}^{Q,\min},g_{i,t}^{Q,\max}]$, $i \in \set{G} \subseteq \set{N}$.
The active and reactive power flows are denoted as $f_{i,t}^P$ and $f_{i,t}^Q$, $i \in \set{E}$, where $i$ is the index of the \emph{downstream} node of edge~$i$, i.e. the node at the receiving end of edge~$i$. Accordingly, the sending node of edge $i$ is  \emph{upstream}. Each edge~$i$ has resistance~$R_i$, reactance~$X_i$ and apparent power flow limit $S_i^{\max}, i \in \set{E}$.

To model the AC power flows we invoke the \emph{LinDistFlow} formulation \cite{Turitsyn_Local_2010}, which  is a lossless approximation of the branch-flow model \cite{farivar2013branch} that allows to account for active and reactive power flows and voltage magnitudes in a computationally tractable manner. This allows to write the AC power flow equations as follows:
\begin{align}
    (d_{i,t}^P - g_{i,t}^P) + \sum_{j \in \set{C}_i} f_{j,t}^P &= f_{{\set{A}_i},t}^P, 
        & \forall t, \forall i \in \set{N} \label{eq:first_lindist_const}\\
    (d_{i,t}^Q - g_{i,t}^Q) + \sum_{j \in \set{C}_i} f_{j,t}^Q &= f_{{\set{A}_i},t}^Q, 
       & \forall t, \forall i \in \set{N}  \label{eq:medium_lindist_const}\\ 
    u_{\set{A}_i,t} - 2(f_{i,t}^P R_i + f_{i,t}^Q X_i) &= u_{i,t}, 
        & \forall t, \forall i \in \set{N}^+ \label{eq:lindist_volt},
\end{align}
where $u_{0,t}$ is assumed to be the fixed voltage value at the root node of the distribution network that can be set to a desired level by the substation transformer. For the sake of simplicity, it is set as $u_{0,t} = \unit[1]{p.u.}, \forall t \in \set{T}$, in the following. 

{\color{black} 
\begin{remark} 
The \textit{LinDistFlow} power flow formulation is common in distribution system analyses because it accurately computes power flows and voltage magnitudes for radial and meshed topologies with a relatively high accuracy, \cite{Turitsyn_Local_2010}.
Further, we note that the parameter estimation method does not depend on a chosen power flow model and other models can be used instead, \cite{DallAnese_Chance_2017}. 
\end{remark}
}

\subsection{Chance-Constrained Power Flow}

The effect of random  variable $\eps_i$ can be incorporated in the power flow equations given in \eqref{eq:first_lindist_const}-\eqref{eq:medium_lindist_const} as follows: 
\begin{align}
    & \bm{d}_{i,t}^P = \bar{d}_{i,t}^P - (x_{i,t}^* + \eps_i)
    \label{eq:first_pf_constraint} \\
    & \bm{d}_{i,t}^Q = \bar{d}_{i,t}^P - (x_{i,t}^* + \eps_i)\gamma_{i,t},
\end{align}
where $x_{i,t}^*$ denotes the  amount of DR capacity desired by the DSO and $\gamma_{i,t}$ is a parameter that relates the active and reactive power values via a given  power factor value. 
The available amount of demand reduction is limited by the nodal demand:
\begin{align}
    \bar{d}_{i,t}^P - x_{i,t}^* \geq 0. \label{eq:dr_limit}
\end{align}
Given random variables  $\bm{d}_{i,t}^{P}$ and $\bm{d}_{i,t}^{Q}$, the DSO schedules its controllable generation resources to balance  the power consumed and produced. Under the assumption that all generation resources follow an  affine control policy \cite{Bienstock_Chance_2014, DallAnese_Chance_2017, dr-ac-ccopf,hassan2018optimal}, the output of each generation resource can be modeled as
\allowdisplaybreaks
\begin{align}
    &\bm{g}_{i,t}^P = g_{i,t}^P - \alpha_{i,t} \sum_{i \in \set{D}} \bm{\epsilon}_{i,t} \\
    &\bm{g}_{i,t}^Q = g_{i,t}^Q - \alpha_{i,t} \sum_{i \in \set{D}} \bm{\epsilon}_{i,t}\gamma_{i,t},
\end{align}
\allowdisplaybreaks[0]
\noindent
{\color{black} 
where  $\alpha_i \geq 0, \forall i \in \set{G}$, is a participation factor of each generator. By enforcing $\sum_{i\in \set{N}} \alpha_{i,t} = 1, \forall t \in \set{T}$, the scheduled balancing capacity is  sufficient to deal with  the assumed deviation from the forecast conditions, \cite{Bienstock_Chance_2014, DallAnese_Chance_2017, dr-ac-ccopf,hassan2018optimal}.
}
If there are no controllable generators, the root bus provides balancing and, therefore, $\alpha_0 = 1$ and $\alpha_i = 0, \forall i \in \set{N}^+$.

Similarly, we can express uncertain AC power flows and voltage magnitudes as a function of random vector~$\eps$: 
\begin{align}
     & \bm{f}_{i,t}^P = f_{i,t}^P + A_{(i,*)}C \eps \\
     & \bm{f}_{i,t}^Q = f_{i,t}^Q + A_{(i,*)}C \gamma_t \eps \\
     & \bm{u}_{i,t} = u_{i,t} - 2 A_{(*,i)}^\top (R AC + X A C\gamma_t ) \eps,
    \label{eq:last_pf_constraint}     
\end{align}
where $A\in\{0,1\}^{n\times m}$ such that $A_{(i,j)}=1$, if line $i$ is part of the path from root to bus $j$, and $A_{(i,j)}=0$ otherwise, and matrix ${C\coloneqq \alpha e^\top - I}$ maps $\eps$ into changes of nodal injections.

Given  \eqref{eq:first_pf_constraint}-\eqref{eq:last_pf_constraint}, we can enforce the following chance constraints on distribution network operations:
\allowdisplaybreaks
\begin{align}
     & \mathbb{P}[\bm{g}_{i,t}^P \leq g_i^{P,max}] \geq 1-\eta_g,  \quad    \forall t \in \set{T},  \forall i \in \set{N} \label{eq:fist_cc}\\
     & \mathbb{P}[\bm{g}_{i,t}^P \geq g_i^{P,min}] \geq 1-\eta_g,  \quad     \forall t \in \set{T},  \forall i \in \set{N} \label{eq:cc_gP_low}\\
     & \mathbb{P}[\bm{g}_{i,t}^Q \leq g_i^{Q,max}] \geq 1-\eta_g,  \quad     \forall t \in \set{T},  \forall i \in \set{N} \label{eq:cc_gQ_up}\\
     & \mathbb{P}[\bm{g}_{i,t}^Q \geq g_i^{Q,min}] \geq 1-\eta_g,  \quad     \forall t \in \set{T},  \forall i \in \set{N} \label{eq:cc_gQ_low}\\
     & \mathbb{P}[\bm{u}_{i,t} \leq u_i^{max}] \geq 1-\eta_v,  \quad     \forall t \in \set{T},  \forall i \in \set{N} \label{eq:upper_voltage}\\
     & \mathbb{P}[\bm{u}_{i,t} \geq u_i^{min}] \geq 1-\eta_v,  \quad     \forall t \in \set{T},  \forall i \in \set{N} \label{eq:lower_voltage} \\ 
     & \mathbb{P}[(\bm{f}_{i,t}^P)^2 + (\bm{f}_{i,t}^Q)^2  \geq (S_i^{\max})^2] \geq 1-\eta_f, \quad     \forall t, \forall i
     \label{eq:last_cc}
\end{align}
\allowdisplaybreaks[0]
where parameters $\eta_g, \eta_v, \eta_f \in (0,\nicefrac{1}{2}]$ are small, non-negative numbers that define the likelihood of constraint violations, \cite{Bienstock_Chance_2014,Summers_Stochastic_2014,dr-ac-ccopf}.
{\color{black} 
Eq. \eqref{eq:fist_cc}-\eqref{eq:last_cc} are individual chance constraints, i.e. they limit the risk of individual constraint violations and allow for fine-tuning robustness of the DRCC-OPF solution against a particular high-risk component. Enforcing individual chance constraints instead of joint chance constraints is appropriate in distribution system analyses due to a  low number of active constraints \cite{baker2017efficient} and makes it possible to avoid seeking unnecessarily conservative  and computationally demanding joint feasibility guarantees, \cite{vrakopoulou2013probabilistic,roald2017chance,alsac1974optimal}.
}

\subsection{Objective Function}
\label{ssec:objective}
The actual real-time outcome of random variables and thus the actual operating cost are unknown \textit{ex-ante}. Therefore, the DSO minimizes the expected value of the following cost:
\begin{align}
    \Eptn[\bm{C}_t] = \Eptn\left[\bm{C}^{\text{(generation)}}_t\right] + \Eptn\left[\bm{C}^{\text{(sale)}}_t\right] + \Eptn\left[\bm{C}^{\text{(DR)}}_t\right]. \label{eq:expected_cost}
\end{align}
\noindent
The production cost of each  resource is  quadratic:
\begin{align}
    c_i(\bm{g}_{i,t}^P) = c_{2i} (\bm{g}_{i,t}^P)^2 + c_{1i} (\bm{g}_{i,t}^P)^2 + c_{01},
\end{align}
so the system-wide generation cost is:
\begin{subequations}
\begin{align}
    \sum_{i \in \set{G}} c_i(\bm{g}^P_i) = c_{2,t} \left(\sum_{i \in \set{D}} \bm{\epsilon}_i\right)^2 + c_{1,t} \sum_{i \in \set{D}}  \bm{\epsilon}_i + c_{0,t}, 
\end{align}
where:
\begin{equation}
    \begin{cases}
        c_{2,t} &\coloneqq \sum_{i \in \set{G}} c_{i2} \alpha_{i,t}^2 \\
        c_{2,t} &\coloneqq \sum_{i \in \set{G}} \left(2 c_{i2} \alpha_{i,t} \bar{g}^P_{i,t} + c_{i1} \alpha_{i,t} \right) \\
        c_{0,t} &\coloneqq \sum_{i \in \set{G}} \left(c_{i2} (\bar{g}^P_{i,t})^2 + c_{i1}\bar{g}^P_{i,t} + c_{i0} \right). 
    \end{cases}
\end{equation}
\end{subequations}
Note that while the cost of distributed generation at nodes $i=1,...,n$ can be assumed fixed over time, cost at the root node ($i=0$) are driven by the wholesale price $\omega_t$ and thus vary over time.
As follows from \eqref{eq:error_definition}, $\Eptn[\eps] = 0$ and  the expected production cost of all generation resources can be computed as:
\allowdisplaybreaks
\begin{align}
    &\Eptn\left[\bm{C}^{\text{(generation)}}_t\right] = \sum_{i \in \set{G}} c_i(g_{i,t}^P) 
        + \sum_{i \in \set{G}}\left(c_{i2} \alpha_i^2 \Var[e^\top \eps] \right).
        \label{eq:expected_costs}
\end{align}%
\allowdisplaybreaks[0]%
Similarly, the lost revenue of not selling $\sum_{i\in\set{N}}x^*_{i,t}$ at retail tariff $\kappa_t$ is computed as:
\begin{align}
    \Eptn[\bm{C}^{\text{(sale)}}_t] = \kappa_t \sum_{i \in \set{N}}x^*_{i,t}
    \label{exp:sale_cost}
\end{align}
where $\kappa$ is a given retail tariff for electricity. 

The last term in \eqref{eq:expected_cost} represents the remuneration that the DSO needs to pay to DR participants, which depends on the desired amount of demand response $x_{i,t}^*$ and price signal $\lambda$. Using \eqref{eq:model_expected_dr} and the desired DR capacity $x_{i,t}^*$, price signal can be computed as follows:
\begin{align}
    \lambda_{i,t} = \frac{x_{i,t}^* - \hat{\beta}_{0,i}^{(t)}}{2\hat{\beta}_{1,i}^{(t)}}
    \label{eq:price_signal_calc}
\end{align}
where we have to make the technical assumption $\hat{\beta}_{1i}^{(t)} \neq 0$. This assumption is not restrictive because estimattions close to zero will lead to prohibitively high price signals, which will lead to the same result as if  true parameter $\beta_{1,i}$ were actually equal to zero (i.e. a node that is insensitive to DR incentive signals).
Accordingly, the last term in \eqref{eq:expected_cost} can be recast as:
\begin{align}
    \Eptn[\bm{C}^{\text{(DR)}}_t] = \sum_{i \in \set{N}} x_{i,t}^*  \frac{x_{i,t}^* - \hat{\beta}_{0,i}^{(t)}}{2\hat{\beta}_{1,i}^{(t)}}.
    \label{eq:exp_dr_cost}
\end{align}

The objective function and chance constraints formulated above give the following AC CC-OPF problem:
\begin{subequations}
\begin{align}
    & \forall t \in \set{T}: \min_{x_{i,t}^*,g_{i,t}^{P},g_{i,t}^{Q},f_{i,t}^{P},f_{i,t}^{Q},\alpha_{i,t}} \Eptn[\bm{C}_t] \\
    s.t. \nonumber\\
    & \eqref{eq:first_pf_constraint} - \eqref{eq:last_pf_constraint}    \qquad \text{: [Power flow equations]} \\
    & \eqref{eq:fist_cc} - \eqref{eq:last_cc}  \qquad \text{: [Chance Constraints]}.
\end{align}
\label{eq:full_cc_model}
\end{subequations}

\subsection{Distributionally Robust Reformulation} \label{sec:dro}
As the uncertainty distribution underlying $\eps$ is unknown \textit{a~priori},  chance constraints in \eqref{eq:fist_cc} - \eqref{eq:last_cc} cannot be reformulated into SOCP constraints as common for various parametric distributions, \cite{Bienstock_Chance_2014}. Consistently  with distribution-free assumptions in price sensitivity learning in Section~\ref{sec:parameter_estimation}, we extend the AC CC-OPF formulation in \eqref{eq:full_cc_model} to a distributionally robust form that eliminates the need in invoking potentially erroneous distribution assumptions. 

Empirical mean and variance values of the residual errors given in  \eqref{eq:emp_mean}-\eqref{eq:emp_var} can be associated with multiple distributions that are collected in  set $\set{P}$. Using set $\set{P}$, the upper voltage chance constraint \eqref{eq:upper_voltage} yields the following distributionally robust chance constraint:
\begin{align}
    \inf_{\mathbb{P} \in \set{P}} \mathbb{P}[\bm{u}_{i,t} \leq u_i^{max}] \geq 1-\eta_v \qquad \forall t \in \set{T}, \forall i \in \set{N}.
    \label{eq:dr_cc_voltageconst}
\end{align}

To reformulate distributionally robust constraint \eqref{eq:dr_cc_voltageconst} in a tractable form, \eqref{eq:last_pf_constraint} is recast as follows:
\begin{align}
    \bm{u}_{i,t} = u_{i,t} + T_i(\alpha) \eps, \label{eq:safety}
\end{align}
where:
\begin{align}
    T_i(\alpha) \coloneqq -2 A_{(*,i)}^\top (R AC + X A C\gamma_t ).
\end{align}
Term  $T_i(\alpha) \eps$ in \eqref{eq:safety} represents the  effect of fluctuations imposed by random vector $\eps$ on the voltage magnitude at node~$i$. These fluctuations must be contained within given voltage limits $u_i^{max}$: 
\begin{align}
    u_i^{max} - s^{v,max}_{i,t} \geq u_{i,t},
    \label{eq:slack_margin}
\end{align}
where $s^{v,max}_{i,t}$ is a slack variable that represents the distance between baseline value $u_{i,t}$ and its limit $u_i^{max}$. Naturally, if  $T_i(\alpha) \epsilon_t \leq s^{v,max}_{i,t} $ holds for a given realization $\epsilon_t$ of $\eps$  then the fluctuations are within the limit. Therefore, \eqref{eq:dr_cc_voltageconst} can be equivalently reformulated as:
\begin{align}
    \inf_{s^{v,max}_{i,t}, \mathbb{P} \in \set{P}}  \mathbb{P} [s^{v,max}_{i,t} \geq T_i(\alpha) \eps] \geq 1 \!-\!\eta_v, \quad \forall t \in \set{T}, \forall i \in \set{N}. \label{eq:dsitro}
\end{align}

The optimal solution of \eqref{eq:dsitro} is the smallest value of slack variable $s^{v,max}_{i,t}$ that ensures that the distributionally robust chance constraints holds with confidence level $1-\eta_v$. This interpretation relates the solution of~\eqref{eq:dsitro} to the concept of \textit{Conditional Value at Risk} (CVaR). Accordingly, the optimal value of $s^{v,max}_{i,t}$ is attained, if \eqref{eq:dsitro} is replaced  by the following set of  matrix inequalities using \cite[Theorem 2.1]{zymler2013distributionally}:
\allowdisplaybreaks
\begin{subequations}
\begin{align}
 \hspace{-3mm}  \forall t \in \set{T}&, \forall i \in \set{N}: \nonumber \\
     & \hspace{-5mm} M^{v,max}_{i,t} \succeq 0   \label{eq:sd_const1} \\
     & \hspace{-5mm}  s^{v,max}_{i,t} + \frac{1}{\eta_v}\scal{\hat{\Omega}^{(t)}, M^{v,max}_{i,t}} \leq 0,  \\
     & \hspace{-5mm} M^{v,max}_{i,t} - \begin{bmatrix}
             0 &  \frac{1}{2}T_i(\alpha)^\top \\
             \frac{1}{2}T_i(\alpha) &  u_{i,t} - u_i^{max} - s^{v,max}_{i,t}
         \end{bmatrix} 
         \succeq 0,   \label{eq:sd_const2}
\end{align}%
\end{subequations}%
\allowdisplaybreaks[0]%
where $s^{v,max}_{i,t}$ and auxiliary matrix $M^{v,max}_{i,t}$ are decision variables and $\hat{\Omega}^{(t)}$ is the second-order moment matrix:
\begin{align}
    \hat{\Omega}^{(t)} \coloneqq \begin{bmatrix}
    \hat{\Sigma}^{(t)} + \hat{\mu}^{(t)} (\hat{\mu}^{(t)})^\top   & \hat{\mu}^{(t)} \\
    (\hat{\mu}^{(t)})^\top  & 1
    \end{bmatrix},
\end{align}
where parameters  $\hat{\mu}^{(t)}$  and $\hat{\Sigma}^{(t)}$ are learned from the LSE as explained in \eqref{eq:emp_mean} and \eqref{eq:emp_var}, respectively. Eq.~\eqref{eq:sd_const1}-\eqref{eq:sd_const2} contain semidefinite constraints that can be solved efficiently by off-the-shelf solvers, e.g. MOSEK \cite{mosek}.
By inferring the security margin of the chance-constraints from the empirical error distribution, the model can not only learn the price sensitivity but also the distribution of the disturbance in an online fashion. This allows both to optimize price signals and to trade-off between larger security margins and higher costs.

The same procedure can be applied to obtain semidefinite reformulation of other chance constraints in \eqref{eq:full_cc_model}. The final result for these reformulations is presented in  Appendix \ref{ax:comlete_set_of_dr_chance_constraints}.

{\color{black} 
\begin{remark}
The proposed approach requires a distributionally robust optimization method to accommodate the mixture of errors in the observable residuals and, thus, the unknown error distribution.
The proposed distributionally robust formulation is independent from the parameter learning process and its conservatism can be tuned via  risk parameter $\eta$.
\end{remark}
}

{\color{black} 
\subsection{Theoretical Regret Analysis}

The anticipated regret associated with the proposed learning approach can be defined as the expected difference between the cost attained with estimated parameters and the cost postulated for true (unknown) parameters. 
As shown in Appendix~\ref{ax:regret_components}, the total regret can be computed as $\zeta(t) = \zeta^{\text{[en]}}(t) + \zeta^{\text{[bal]}}(t)$, where the expected regret due to the cost of energy procurement is:
\allowdisplaybreaks
\begin{equation}
\begin{split}
\zeta^{\text{[en]}}(t) = \sum_{i\in\set{N}}\! \underbrace{\left(\!(\frac{1}{2\hat{\beta}_{1,i}^{(t)}} - \frac{1}{2\beta_{1,i}})(x_{i,t}^*)^2\!-\! (\frac{\hat{\beta}_{0,i}^{(t)}}{2\hat{\beta}_{1,i}^{(t)}}\! -\! \frac{\beta_{0,i}}{2\beta_{1,i}}) x_{i,t}^*\!\right)}_{\coloneqq \zeta_i^{\text{[en]}}(t)}
\end{split}%
\label{eq:regret_energy_provision}%
\end{equation}%
\allowdisplaybreaks[0]%
\noindent
and the expected regret due to the balancing cost is: 
\allowdisplaybreaks
\begin{equation}
\begin{split}
   \zeta^{\text{[bal]}}(t) = \sum_{i\in\set{N}} 
 \underbrace{\left( \alpha_i^2 c_{2i} e^{\!\top}(\hat{\Sigma}^{(t)} - \Sigma )e \right)}_{\coloneqq\zeta_i^{\text{[bal]}}(t)},
\end{split}%
\label{eq:regret_balancing}%
\end{equation}%
\allowdisplaybreaks[0]%
where $\Sigma = \diag([\sigma_1^2,...,\sigma_n^2])$.

Regret component $\zeta^{\text{[en]}}$ in \eqref{eq:regret_energy_provision} depends on the parameter estimation error, i.e. the discrepancy between $\hat{\beta}_i^{(t)}$ and $\beta_i$, and the amount of desired demand response $x_{i,t}^*$ at each node.
On the other hand, regret component $\zeta^{\text{[bal]}}$ is driven by the empirical variance of the desired demand response ($\hat{\Sigma}^{(t)}$), see \eqref{eq:emp_var}.
To further analyze $\zeta^{\text{[en]}}$ and $\zeta^{\text{[bal]}}$, we first consider the optimality condition for $x_{i,t}^*$:
\begin{prop}
\label{prop:optimal_x}
Consider the OPF problem given by \eqref{eq:full_cc_model}.
Let $\pi_{i,t}^p$ and $\pi_{i,t}^q$ be the Lagrangian dual multipliers of the active and reactive nodal power balances at node $i$ and time $t$ in \eqref{eq:first_lindist_const} and \eqref{eq:medium_lindist_const}.
Then the optimal desired DR at each node $i$ is given as
\begin{equation}
    x^*_{i,t} = \hat{\beta}_{1,i}^{(t)} (\pi_{i,t}^p + \gamma_i \pi_{i,t}^q - \kappa_t) + \hat{\beta}_{0,i}^{(t)}.
    \label{eq:optimal_x_inprop}
\end{equation}
\end{prop}
\myproofstart
The first-order optimality condition of $x_{i,t}^*$ in \eqref{eq:full_cc_model} is:
\begin{equation}
    \pi_{i,t}^p + \gamma_i \pi_{i,t}^q = \frac{1}{\hat{\beta}_{1,i}^{(t)}} x_{i,t}^* - \frac{\hat{\beta}_{i,t}^{(t)}}{\hat{\beta}_{1,i}^{(t)}} + \kappa_t.
    \label{eq:kkt_x}
\end{equation}
Re-arranging \eqref{eq:kkt_x} immediately leads to \eqref{eq:optimal_x_inprop}.
\myproofend

It follows from Proposition~\ref{prop:optimal_x} and \eqref{eq:price_signal_calc} that the optimal price signal to achieve optimal $x_{i,t}^*$ is given as:
\begin{equation}
\lambda_{i,t}^* = \pi_{i,t}^p + \gamma_i \pi_{i,t}^q - \kappa_t. \label{eq:optimal_price_signal}
\end{equation}
Since $ \lambda_{i,t}^* \geq 0$ by definition, any node $i$ receives a non-zero price signal in $t$ only if \mbox{$\pi_{i,t}^p + \gamma_i \pi_{i,t}^q > \kappa_t$}.
Next, we analyze the convergence of the parameter estimation error using the results of Proposition~\ref{prop:optimal_x}.
\begin{prop}
\label{prop:convergence_of_estimation_error}
Let $\lambda_{i,t}^*$ be the broadcast price signal at node~$i$ and time~$t$, and $\pi_{i,t}^p$, $\pi_{i,t}^q$ the Lagrangian dual multipliers of the active and reactive nodal power balances at node $i$ and time $t$ of \eqref{eq:full_cc_model}, and let $B_t \coloneqq \hat{\beta}_i^{(t)}-\beta_i$ be the parameter estimation error. 
If the broadcast price is given by
\begin{align}
    \lambda_{i,t}^* = \max(\pi_{i,t}^p + \gamma_i \pi_{i,t}^q - \kappa_t, 0),
    \label{eq:truncated_price_signals}
\end{align}
then parameter estimation error $B_t$ converges to zero for all $t$, where $\lambda_{i,t}>0$.
\end{prop}
\myproofstart
Consider the parameter estimation error as:
\begin{align}
B_t = 
\begin{bmatrix}
    \hat{\beta}_{1,i}^{(t)} - \beta_{1,i} \\ \hat{\beta}_{0,i}^{(t)}  - \beta_{0,i}
\end{bmatrix} = F_{i,t}^{-1} \left( \sum_{\tau=1}^{t-1} \begin{bmatrix} \lambda_{i,\tau} \\ 1 \end{bmatrix} \hat{\epsilon}_{\tau}^{(t)} \right), 
\label{eq:parameter_est_error}
\end{align}
where:
\begin{align}
    F_{i,t} = \begin{bmatrix} \sum_{\tau=1}^{t-i} \lambda_{i,\tau}^2 & \sum_{\tau=1}^{t-i} \lambda_{i,\tau} \\ \sum_{\tau=1}^{t-i} \lambda_{i,\tau} & (t-1) \end{bmatrix},
\end{align}
is the Fisher information of node $i$ at time $t$, \cite{keskin2014dynamic, den2013simultaneously}.
It follows from \eqref{eq:parameter_est_error} that the parameter estimation error converges to zero, if the minimum eigenvalue of $F_{i,t}$ increases unbounded over time, \cite{keskin2014dynamic, den2013simultaneously}. 
Recalling Lemma~2 of \cite{keskin2014dynamic}, the minimum eigenvalue of $F_{i,t}$ is bounded from below as:
\begin{equation}
\begin{split}
    L_{i,t} &= \sum_{\tau=1}^{t-i} (\lambda_{i,\tau} - \bar{\lambda_{i,t}})^2 \\
    &= (t-2)\Var([\lambda_{i,\tau}, \tau=\{1,...,t-1\}]).
    \label{eq:squared_price_deviatons}
\end{split}
\end{equation}
Eq. \eqref{eq:squared_price_deviatons} shows that $L_{i,t}$ increases over time if the variance of the broadcast price signals $\Var([\lambda_{i,\tau}, \tau=\{1,...,t-1\}])$ does not converge to zero. 
Under the non-restrictive assumption that the root node electricity price $\omega_t$ changes over time, i.e. is different for different $t$, $\pi_{i,t}^p$ and $\pi_{i,t}^q$ are similarly changing over time due to their dependency on the cost of energy provision and the active network constraints, \cite{Papavasiliou_Analysis_2017}.
It follows from the relation between $\lambda_{i,t}^*$ and $\pi_{i,t}^p$, $\pi_{i,t}^q$ given by \eqref{eq:truncated_price_signals} that  $\lambda_{i,t}^* \neq 0$ will not be uniform across different $t$. 
Thus, if $\set{T}_i^+ \subseteq \{1,...,t-1\}$ is the set of timesteps with $\lambda_{i,t} > 0$, then 
\begin{align}
    \Var([\lambda_{i,\tau}^*, \tau\in\set{T}_i^+]) > 0. 
    \label{eq:bounded_lambda_var}
\end{align}
It follows from \eqref{eq:bounded_lambda_var} and \cite[Lemma 2]{keskin2014dynamic} that parameter estimation error $B_t$ given in \eqref{eq:parameter_est_error} converges to zero over time. 
\myproofend

The results of Propositions~\ref{prop:optimal_x} and \ref{prop:convergence_of_estimation_error} lead to the following result on the convergence of regret:
\begin{prop}
Let regret the regret be $\zeta(t) = \zeta^{\text{[en]}}(t) + \zeta^{\text{[bal]}}(t)$, where $\zeta^{\text{[en]}}(t)$ and $\zeta^{\text{[bal]}}(t)$ are given by \eqref{eq:regret_energy_provision} and \eqref{eq:regret_balancing}.
If at every time step $t$ the price signal is chosen as \eqref{eq:truncated_price_signals}, then aggregated regret $\frac{1}{t}\sum_{\tau=1}^{t-i}\zeta(t)$
is sublinear over $t$.
\end{prop}
\myproofstart
First, consider:
\begin{align}
    \sum_{\tau}^{t-1} \zeta^{\text{[en]}}(\tau) 
     = \sum_{i\in\set{N}} \left(\sum_{\tau\in\set{T}_i^+} \zeta_i^{\text{[en]}}(\tau) + \sum_{\set{T}_i^0} \zeta_i^{\tau\in\text{[en]}}(\tau)\right).
    \label{eq:agg_regret}
\end{align}
where $\set{T}^0 = \{1,...,t-1\}\setminus\set{T}_i^+$ so that $\set{T}_i^+ \cup \set{T}^0 = \{1,...,t-1\}$ and $\set{T}_i^+ \cap \set{T}^0 = \emptyset$.
As shown in Proposition~\ref{prop:convergence_of_estimation_error}, at every time step $t$  with  $\lambda_{i,t}^* \neq0$ the parameter estimation error at this node decreases on average.
Therefore, as follows from \eqref{eq:regret_energy_provision}, the regret contribution of this node decreases on average as well. 
On the other hand, any node $i$ where $\pi_{i,t}^p + \gamma_i \pi_{i,t}^q < \kappa_t$ and thus $\lambda_{i,t}^* = x_{i,t}^* = 0$, has a zero contribution to $\zeta^{\text{[en]}}$ as per \eqref{eq:regret_energy_provision} so that $\sum_{\tau\in\set{T}_i^0}\zeta_i^{\text{[en]}}(\tau)= 0$.

Next, we consider $\zeta^{\text{[bal]}}$ in \eqref{eq:regret_balancing}.
Unlike for $\zeta^{\text{[en]}}$, information on the random error is acquired at every time step, even if desired DR participation $x^*_{i,t} = 0$, which leads to the convergence of $\zeta^{\text{[bal]}}$.
The convergence of the individual regret components $\lim_{t\rightarrow\infty} \zeta^{\text{[en]}}(t) = \lim_{t\rightarrow\infty} \zeta^{\text{[bal]}}(t) = 0$ leads to  
\begin{equation}
    \frac{1}{t}\sum_{\tau=1}^{t-i}\left( \zeta^{\text{[en]}}(\tau) + \zeta^{\text{[bal]}}(\tau)\right) = \Theta(\log(t)),
\end{equation}
i.e. a sublinear trajectory of the aggregated regret over time.
\myproofend

Note that if the network is unconstrained (i.e. no line or voltage constraint is binding), then $\pi_{i,t}^p = \pi_{t}^p, \forall i\in\set{N}$, and $\pi_{i,t}^q = \pi_{t}^q, \forall i\in\set{N}$, resulting in $\lambda_{i,t}^* = \lambda_t^*, \forall i\in\set{N}$, which leads to the similar regret guarantees as in \cite{Li_A_2017}, where no physical network constraints are modeled.
}

\section{Case Study}

\begin{figure}[t]
\centering
\includegraphics[width=0.4\linewidth]{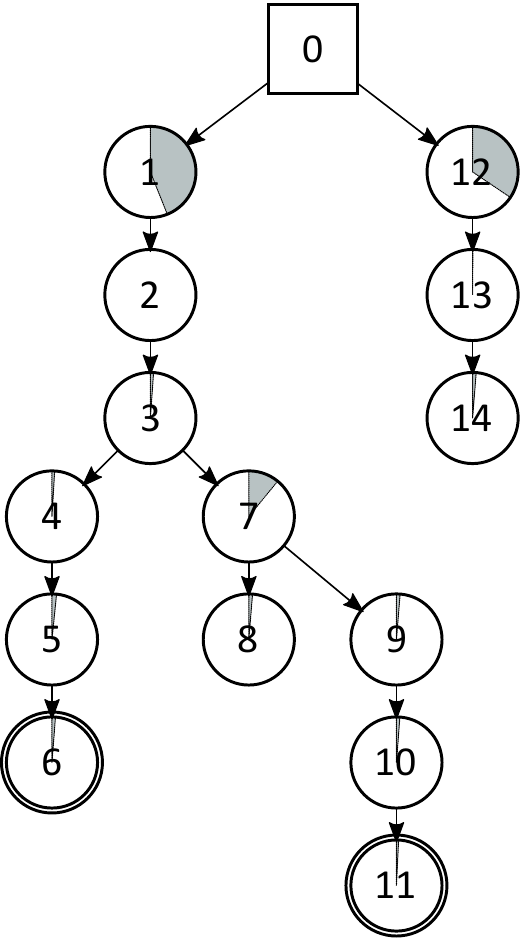}
\caption{ \small The 15-node test system from \cite{Papavasiliou_Analysis_2017}, where the square root node (substation) and the double contour nodes denotes controllable resources. At each node, the filled ratio of the circle indicates the share of the node in the total forecast demand.}
\label{fig:15bus_system}
\end{figure}

{\color{black} 
Fig.~\ref{fig:15bus_system} illustrates the 15-node test system from \cite{Papavasiliou_Analysis_2017} used in the case study with two controllable generators added to nodes 6 and 11. 
Each generator has a linear cost curve with $c_{i,1}=\$\unit[10]{/MWh}$ and  $g^{P,max}_i =\unit[0.8]{MW}$. The time horizon is given by 500 hourly intervals, i.e. $\set{T}=\{1,2,...,500\}$. At each interval, the cost of electricity at the root node is sampled from the range between $\$\unit[30]{/MWh}$ and $\$\unit[200]{/MWh}$ using a uniform distribution. The retail tariff is set to $\kappa_t = \$ 25/MWh, \forall t$. The desired likelihood of chance constraint violations is $\eta_v = \eta_g = 0.1$. 
We use the active and reactive demand from \cite{Papavasiliou_Analysis_2017} as the forecasted baseline and the simulated reaction of the DR participants is generated from the DR model set to the following parameters: $\beta_{1,i} = \unit[\nicefrac{1}{150}]{MWh\$^{-1}}, \forall i \in \set{N}^+$, $\beta_{0,i} = 0, \forall i \in \set{N}^+$, and $\sigma_i = {0.1\bar{d}^P_{i,t}}, \forall i \in \set{N}^+,$ with no correlation among the nodes. Those are the parameters that the model needs to learn over time.
}

To evaluate the effectiveness of the proposed learning procedure, the following four cases representing different levels of information available to the DSO are compared:

\begin{itemize}
    \item \textit{Fully oracle}: The DSO uses the true values of $\beta_i$ and $\Omega$.
    \item \textit{$\beta_i$-oracle}: The DSO uses the true values of $\beta_i$, but $\Omega$ is unavailable and, therefore, $\hat{\Omega}$  needs to be learned.
    \item \textit{$\Omega$-oracle}: The DSO uses the true values of  $\Omega$, but  $\beta_i$ is unavailable and, therefore,  $\hat{\beta}_i$  needs to be learned.
    \item \textit{Fully oblivious}: The DSO must learn both  $\hat{\Omega}$  and $\hat{\beta}_i$.  
\end{itemize}

Additionally, each of the cases above is analyzed for different sets of network constraints in the distribution system:
\begin{itemize}
  \item  \textit{No network}: The network limits in \eqref{eq:upper_voltage}-\eqref{eq:last_cc} are ignored.
  \item \textit{Flow-constrained}: The apparent power flow limit in \eqref{eq:last_cc} is enforced.
  \item \textit{Voltage-constrained}: The voltage limits in \eqref{eq:upper_voltage}-\eqref{eq:lower_voltage} are enforced.
  \item \textit{Fully constrained}: The network limits in \eqref{eq:upper_voltage}-\eqref{eq:last_cc} are enforced.
\end{itemize}

All models in the case study are implemented using the \emph{Julia JuMP} package~\cite{jump}. The code and input data can be downloaded from~\cite{drlearning_code}.

\subsection{DR Learning}

\begin{table}[t!]
\centering
\caption{\small Relative optimal DR usage ($x_{i,t}^*/\sum_i \bar{d}_{i,t}$): (a):~Maximum relative optimal DR, (b):~Median relative optimal DR, (c):~Minimum relative optimal DR, (d):~Median of relative optimal DER utilization, all in \%.}
\label{tab:rel_dr_utilization}
\begin{tabular}{c l|l|l|l|l}
\toprule
  &   & Oracle & $\beta$-oracle & $\Omega$-oracle & Oblivious \\
\midrule
\multirow{4}{*}{\shortstack{No\\Network}} 
                            & (A) & 11.40 & 11.40 & 11.40 & 11.40 \\
                            & (B) & 11.40 & 11.40 & 11.40 & 11.40 \\
                            & (C) & 9.312 & 9.312 & 9.318 & 9.439 \\
                            & (D)  & 100.0 & 100.0 & 100.0 & 100.0 \\
\midrule
\multirow{4}{*}{\shortstack{Only\\Flows}} 
                    & (A) & 67.28 & 67.28 & 69.76 & 69.76 \\
                    & (B) & 40.32 & 40.31 & 40.31 & 40.32 \\
                    & (C) & 5.091 & 5.187 & 5.202 & 5.069 \\
                    & (D)  & 34.16 & 34.16 & 34.17 & 34.15   \\                        
\midrule
\multirow{4}{*}{\shortstack{Only\\Voltage}} &  (A) & 42.06 & 42.78 & 42.09 & 42.74 \\
                                      &  (B) & 42.02 & 41.91 & 41.85 & 41.19 \\
                                      &  (C) & 0.0   & 0.0 & 0.0   & 0.0   \\
                                      &  (D)  & 65.4 & 64.86 & 65.4 & 64.86   \\  
\midrule
\multirow{4}{*}{\shortstack{Fully\\Constrained}}
                                       & (A) &  52.8 & 67.37 & 52.86 &  60.50 \\
                                       & (B) & 40.2 & 40.25 & 40.24 & 40.24 \\
                                       & (C) & 5.04 & 5.042   & 0.0   & 0.0   \\
                                       & (D)  & 34.04 & 34.07 & 34.05 & 34.05  \\  
\bottomrule
\end{tabular}
\end{table}

\textit{1) Optimal DR usage:} Table \ref{tab:rel_dr_utilization} compares the optimal usage of DR resources for different learning cases and sets of network constraints in terms of the total DR amount exercised relative to the total demand in the system, i.e. $x_{i,t}^*/\sum_i \bar{d}_{i,t}, \forall t \in \set{T}$. The case with no network limits enforced leads to a significantly lower usage of DR resources since the DSO can take advantage of the two controllable DERs at node 6 and 11 with production costs lower than the supply from the root node.

However, when the network limits are imposed, the dispatch of DERs becomes more constrained and, therefore, the DSO elects to exercise more DR resources. The usage of DR resources is more affected by voltage limits than by power flow limits due to two factors. First, the distribution systems are typically voltage constrained rather than power flow constrained. Second, as it can also be seen in Table \ref{tab:rel_dr_utilization},  power flow limits prevent the use of controllable DERs by roughly a factor of two relative to the voltage limits. 
{\color{black} 
Notably, the fully constrained case does not necessarily lead to the maximum DR utilization relative to other less constrained cases. 
This result defies the intuition that a more constrained case would require more flexibility.
However, the cost of exercising DR flexibility appear suboptimal in our simulations as network limits affect DR deliverability and  more cost-effective resources are available. 
}

The effect of parameter learning on the optimal usage of DER resources observed in Table~\ref{tab:rel_dr_utilization} is two-fold. First, as the DSO becomes more oblivious to characteristics of DR resources, DR utilization increases relative to the oracle case, while the use of controllable DERs remains nearly the same. Thus, due to a lack of oracular knowledge about DR resources, the DSO is forced to overuse its available DR resources to meet the system-wide demand and avoid violating network limits. Second, as network operations become more restrictive, the difference in the amounts  of DR resources used  in the  fully oracle and fully oblivious cases increases. 

\begin{figure}[t!]
\vspace{-10mm}
\centering
\includegraphics[width=0.95\linewidth]{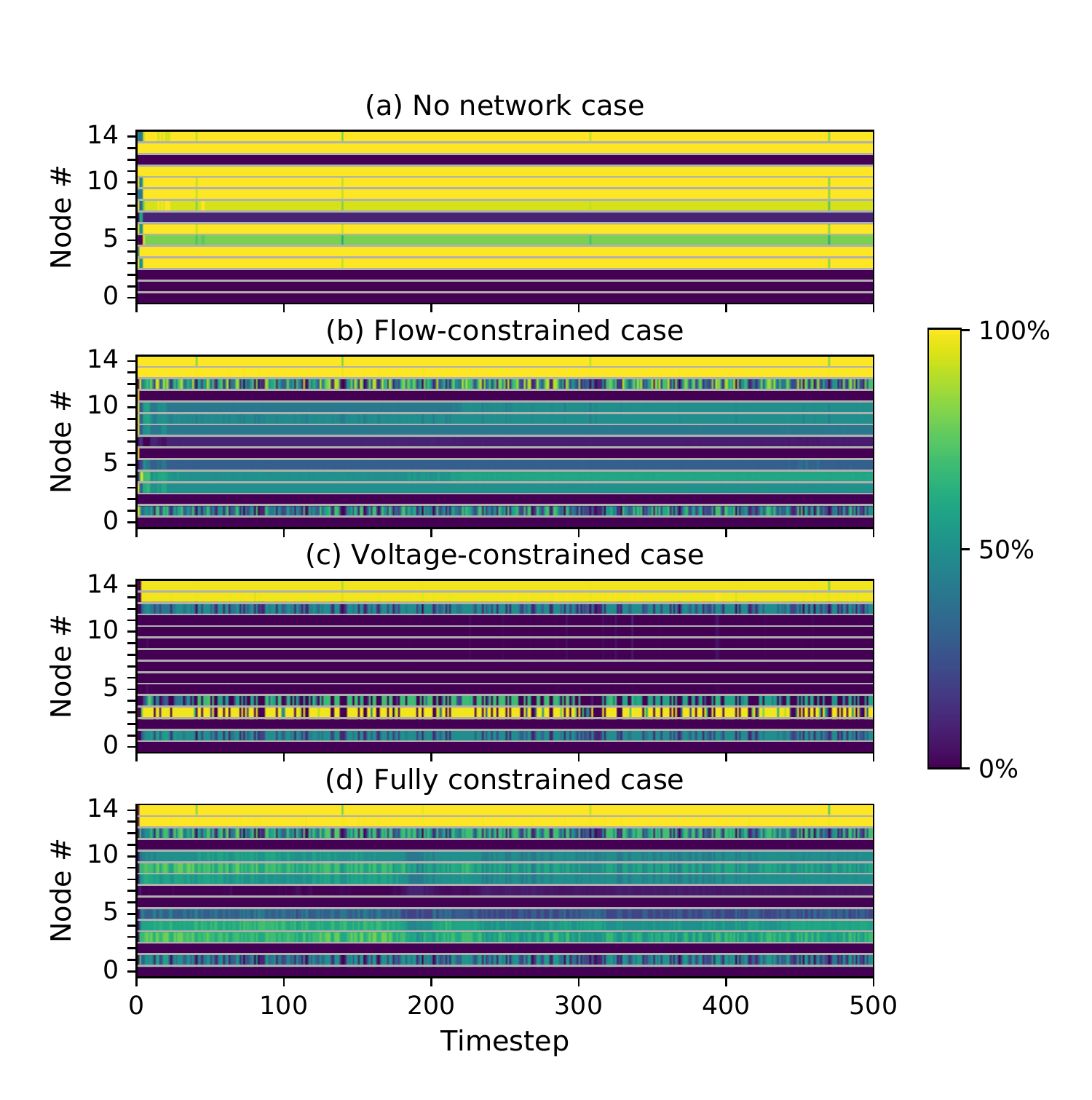}
\caption{ \small Optimal DR usage at nodes relative to the nodal forecast demand, i.e. $x_{i,t}^*/\bar{d}_{i,t}$ .}
\vspace{-5mm}
\label{fig:x_at_bus}
\end{figure}

\begin{figure}[b!]
\vspace{-5mm}
\centering
\includegraphics[width=0.95\linewidth]{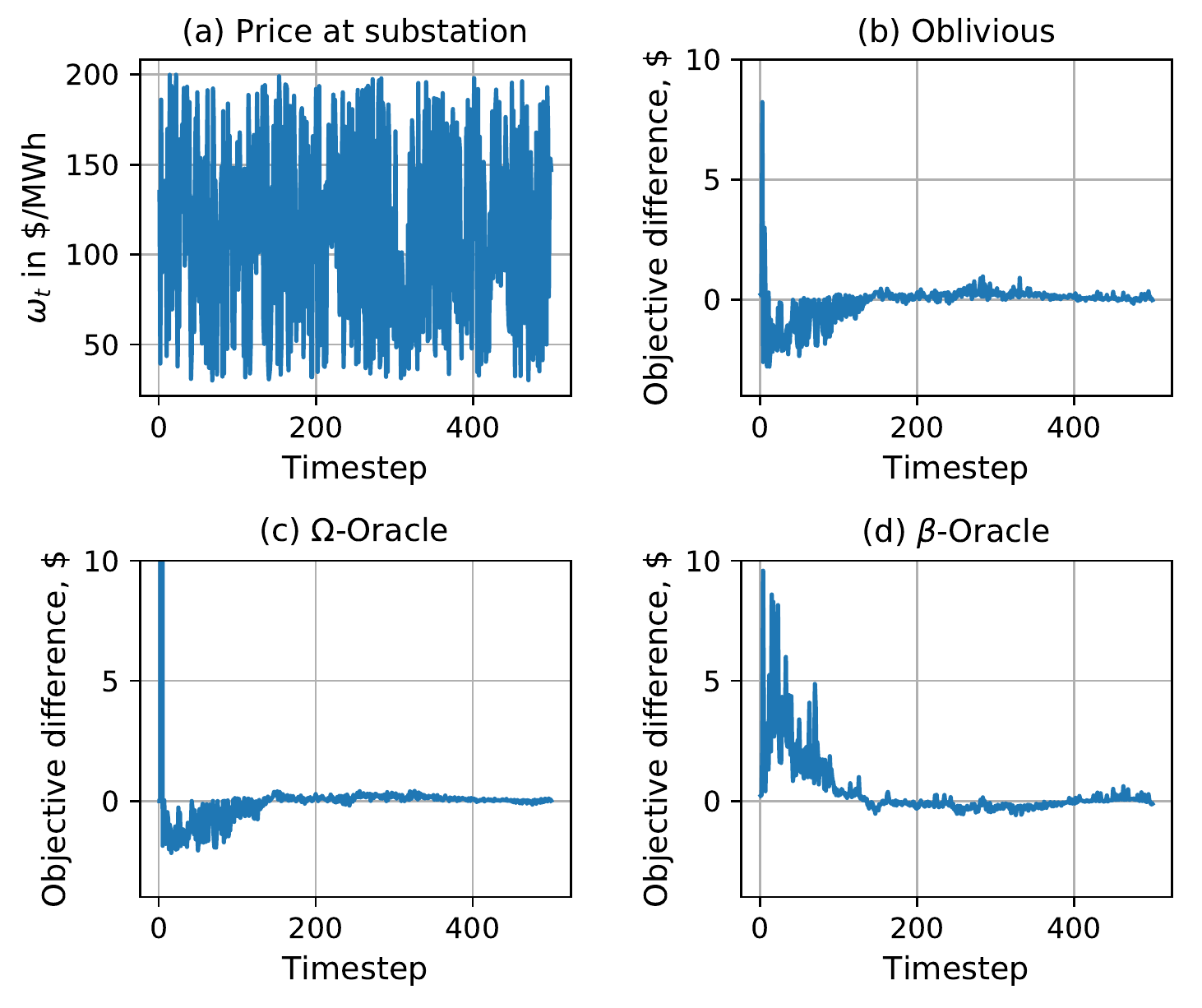}
\caption{ \small (a) Randomly sampled energy price at the root node (substation). (b)-(d) Difference in the DSO objective function between the oracular and non-oracular  cases.}
\label{fig:oracle_diff}
\end{figure}

The aggregated DR usage in the fully oblivious case in Table~\ref{tab:rel_dr_utilization} are itemized for each node and each time interval in {\color{black} Fig.}~\ref{fig:x_at_bus}. While the median aggregated utilization of DR resources reported in Table~\ref{tab:rel_dr_utilization} is roughly the same for all network-constrained cases,  the nodal distribution shown in {\color{black} Fig.}~\ref{fig:x_at_bus} is differently affected by limits imposed. This empirical evidence suggests that tighter network limits forces the DSO to use the DR resources more uniformly across the  system.

\textit{2) Parameter Learning:} Consistently with the cases presented in {\color{black} Fig.}~\ref{fig:x_at_bus}, this section discusses the effect of learning on the DSO objective and presents the outcomes of price learning.  {\color{black} Fig.}~\ref{fig:oracle_diff} compares the DSO objective in the three non-fully-oracular cases, in which some information about DR resources is oblivious,  and the fully oracular case under randomly sampled prices at the root node. As the learning progresses, the accuracy of parameters available to the DSO increases, which reduces the difference between the objective in the oracular and non-oracular cases. This improvement in accuracy is insensitive to the substation price, which indicates the robustness of the proposed learning approach. Among the three cases with non-oracular information, there is no significant difference in convergence.

\begin{figure}[t!]
\centering
\includegraphics[width=0.9\linewidth]{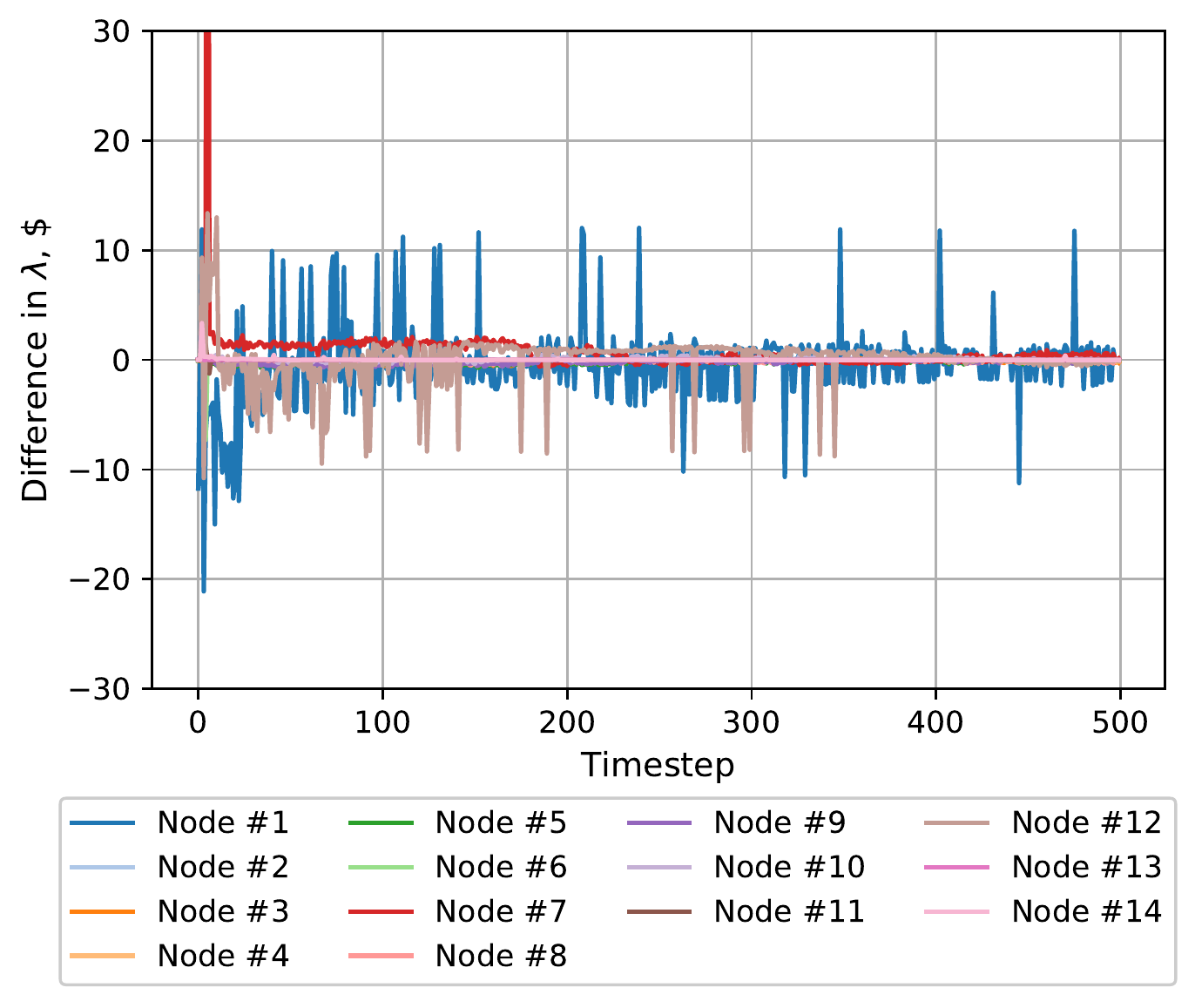}
\caption{Difference between price signals ($\lambda$) obtained in the oracle case and the oblivious case with fully constrained network.}
\label{fig:price_signal_comparison}
\end{figure}

Similarly to the DSO objective, price signals  produced by the proposed learning approach in all non-oracular cases  with all network limits enforced converge to the oracular values, as shown in {\color{black} Fig.}~\ref{fig:price_signal_comparison}. Notably that price signals for all nodes but nodes 1 and 12 converge fairly quickly.
{\color{black} 
The price spikes observed at these two nodes are explained by two factors. First, these nodes carry 75\% of the total load, see Fig.~\ref{fig:15bus_system}, which exacerbates the absolute price difference in Fig.~\ref{fig:price_signal_comparison}  even for small  parameter estimation errors. Second, these two nodes are adjacent to the root node of the distribution system, which amplifies  spikes in the price at the root node, see randomly  generated samples in Fig.~\ref{fig:oracle_diff}(a).  However, the frequency of price spikes at nodes 1 and 12  gradually reduces as the learning procedure progresses. 
}

\subsection{Empirical Analysis of Learning Errors}

In the non-oracular cases, the learning errors steams from the uncertainty~$\eps$ and misestimation of $\hat{\beta}$ and  $\hat \Omega$.  To isolate the effect of misestimated parameters $\hat{\beta}$ and $\hat \Omega$ from~$\eps$, we compute the difference between the expected DSO objective and the observed DSO objective in each case, i.e. $\Delta_t^{[\cdot]} = C_t^{[\cdot]} - \Eptn[C^{[\cdot]}_t],$ where $[\cdot]$ denotes the oracular, $\beta$-oracular, $\Omega$-oracular and oblivious cases, respectively. Since in the oracular case the error inflicted by parameter learning is null by definition, we obtain $\Delta_t^{[\text{oracle}]} = \Delta_t^{[\epsilon]}$, which is the error inflicted by the uncontrollable disturbance $\eps$ in Eq.~\eqref{eq:resulting_demand}. This error is the same in the oracular and non-oracular cases and, therefore, the learning error in the three  non-oracular cases can be recovered as  $
    \Delta_t^{\text{[$\beta$-learning]}} = \Delta_t^{\text{[$\Omega$-oracle]}} - \Delta_t^{\text{[$\epsilon$]}}$, $
    \Delta_t^{\text{[$\Omega$-learning]}} = \Delta_t^{\text{[$\beta$-oracle]}} - \Delta_t^{(\text{[$\epsilon$]})}$ and $
    \Delta_t^{\text{[learning]}} = \Delta_t^{\text{[oblivious]}} - \Delta_t^{\text{[$\epsilon$]}}$, respectively. 

\begin{figure}[!t]
\centering
\includegraphics[width=0.95\linewidth]{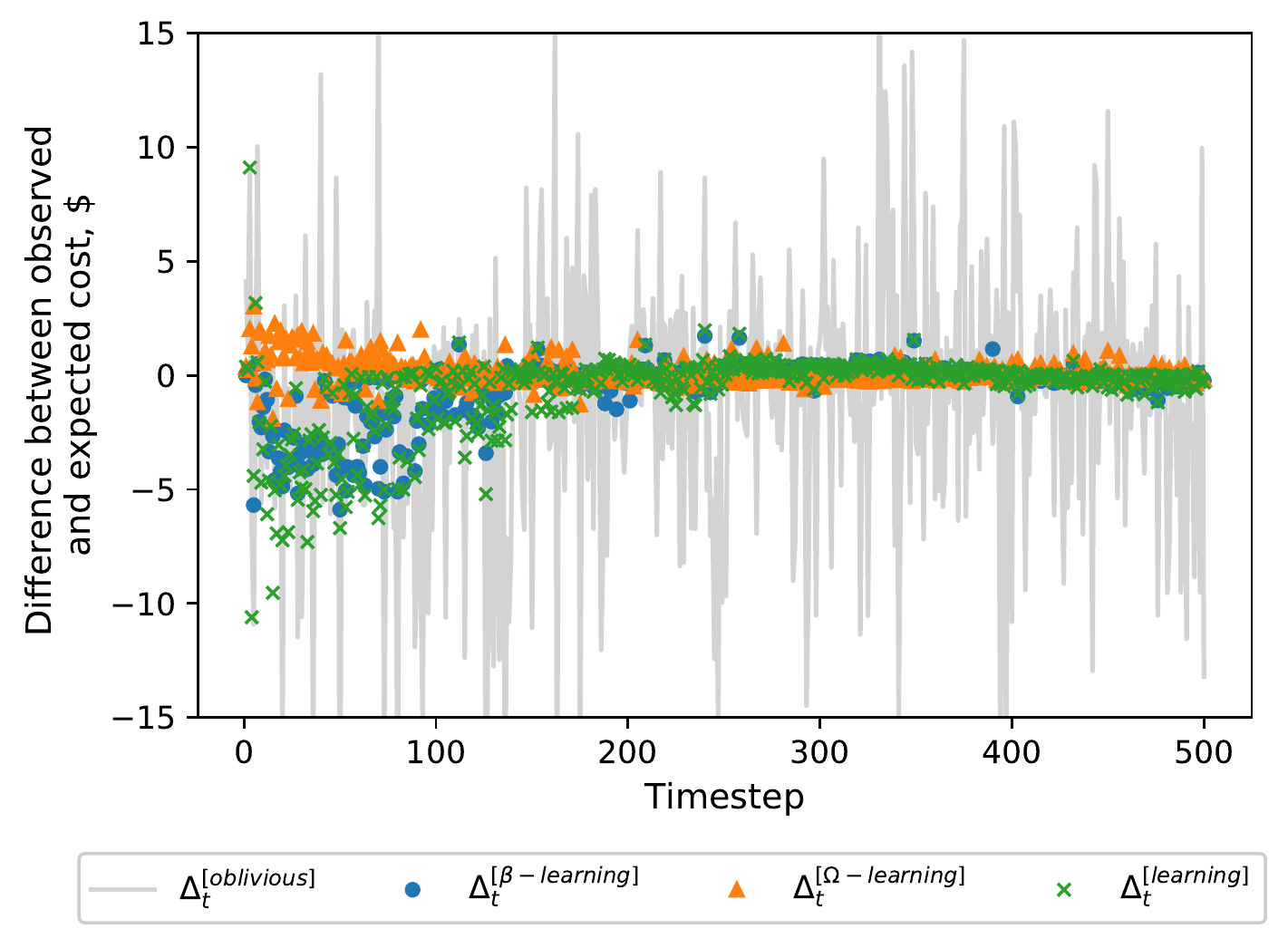}
\caption{Empirical analysis of learning errors for the expected and observed DSO objectives.}
\label{fig:error_comparison}
\end{figure}

{\color{black} Fig.}~\ref{fig:error_comparison} itemizes the learning errors computed as explained above for the cases considered in {\color{black} Fig.}~\ref{fig:oracle_diff}. In all cases, the systematic errors $\Delta_t^{\text{[$\beta$-learning]}}$, $\Delta_t^{\text{[$\Omega$-learning]}}$ and $\Delta_t^{\text{[learning]}}$ converge to zero as the learning progresses. This result demonstrates that the misestimation errors induced by the learning approach can be overcome if sufficient data sets are available.

\begin{figure}[!t]
\centering
\includegraphics[width=0.95\linewidth]{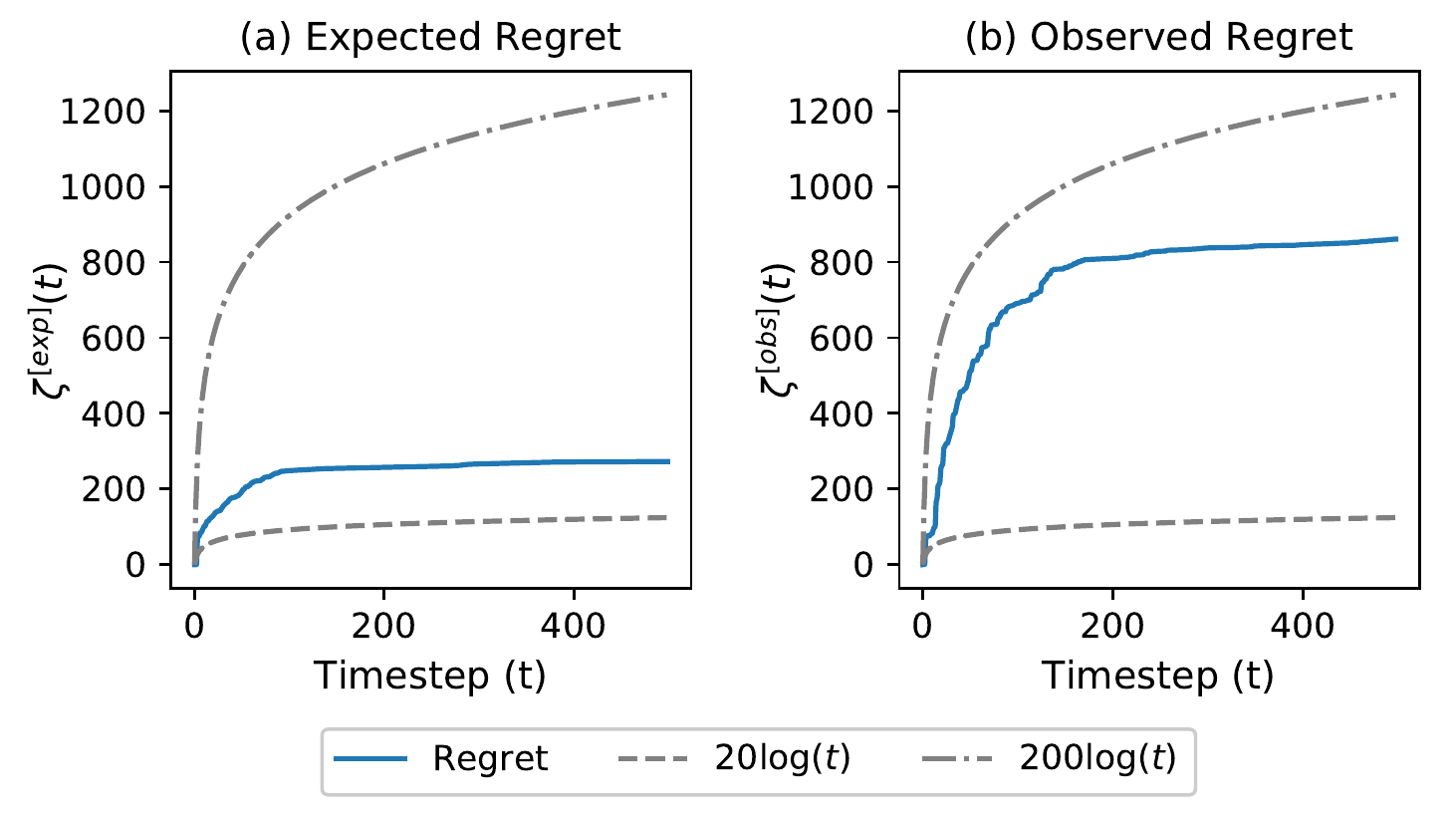}
\caption{Analyses of the expected and observed regret shown within a logarithmic envelope.}
\label{fig:numerical_regret}
\end{figure}

{\color{black} 
\subsection{Experimental Regret Analysis}

Analysis of regret, i.e. the difference between the decision of the oblivious model and the oracle (perfect foresight) model, allows assessment of the performance of the learning process.
We define two regret metrics similar to \cite{Li_A_2017}.
First, the \textit{expected regret} defines the difference between the objective values of the oblivious and oracle models:
\begin{align} \label{eq:exp}
    \zeta^{[\text{exp}]}(t) \coloneqq \sum_{\tau=1}^t \left(\Eptn[\bm{C}_\tau]^{[\text{oblivious}]} - \Eptn[\bm{C}_\tau]^{[\text{oracle}]}\right)^2.
\end{align}
Second, we compute the \textit{observed regret} as the difference between the objective functions of the oblivious and oracle cases after observing the true outcome at each time step: 
\begin{align} \label{eq:obs}
    \zeta^{[\text{obs}]}(t) \coloneqq \sum_{\tau=1}^t \left(C_\tau^{[\text{oblivious}]} - C_\tau^{[\text{oracle}]}\right)^2.
\end{align}

Using \eqref{eq:exp} and \eqref{eq:obs}, we seek a sublinear and  asymptotically zero regret, i.e. $\lim_{t\rightarrow \infty} \nicefrac{\zeta(t)}{t} = 0$, \cite{Khezeli_Risk_2017, den2013simultaneously}. Fig.~\ref{fig:numerical_regret} illustrates the evolution of the expected and observed regret metrics. 
Although the absolute regret value increases as the learning progresses, both regret metrics exhibit a logarithmic trend with the required rate of saturation of $\nicefrac{1}{t}$, as shown by the logarithmic envelope in Fig.~\ref{fig:numerical_regret}.
Note that the scale of the envelope ($20\log(t)$, $200\log(t)$) has been chosen to fit the scale of the shown regret. 
This shows that the regret increment at each time step is on average smaller than at the previous time step, which indicates learning progress at each step. 
The same trend is observed for the evolution of the moments of the residual error, where the difference between the mean and variance in the oblivious and oracle cases reduces as the learning time increases. 
Fig.~\ref{fig:statistics_development} illustrates this evolution for node~10, which was selected since our experiments show that the optimal DR participation at this node has the least sensitivity to the price volatility at the substation. Despite this low sensitivity, we observe that the parameter estimates at node 10 converge. We observe similar convergence trends  at the other nodes of the system.

}

{\color{black}

\begin{figure}
\centering
\includegraphics[width=0.95\linewidth]{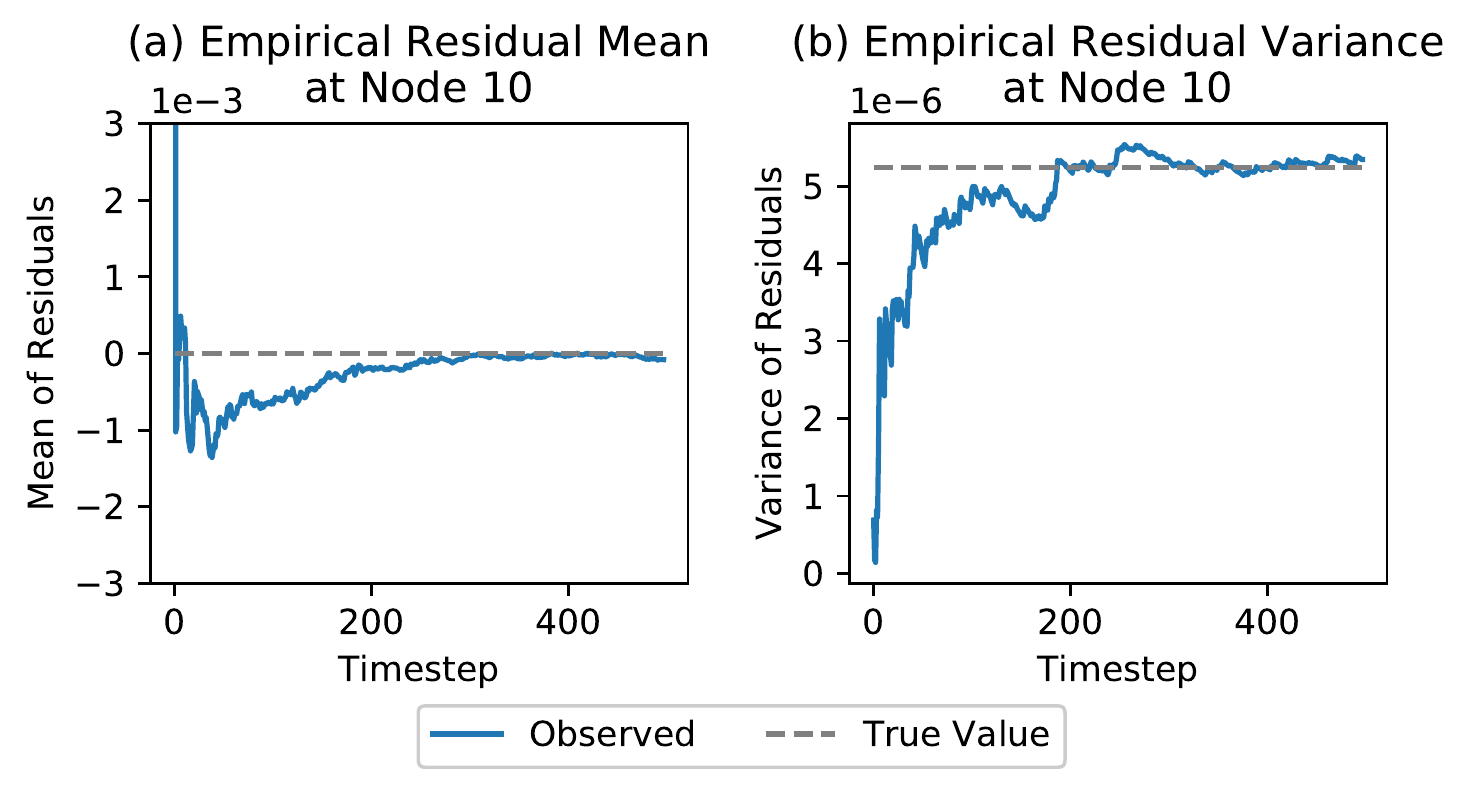}
\caption{Development of empirical mean and variance at node~$10$.}
\label{fig:statistics_development}
\end{figure}

}

\begin{figure}
\centering
\includegraphics[width=0.7\linewidth]{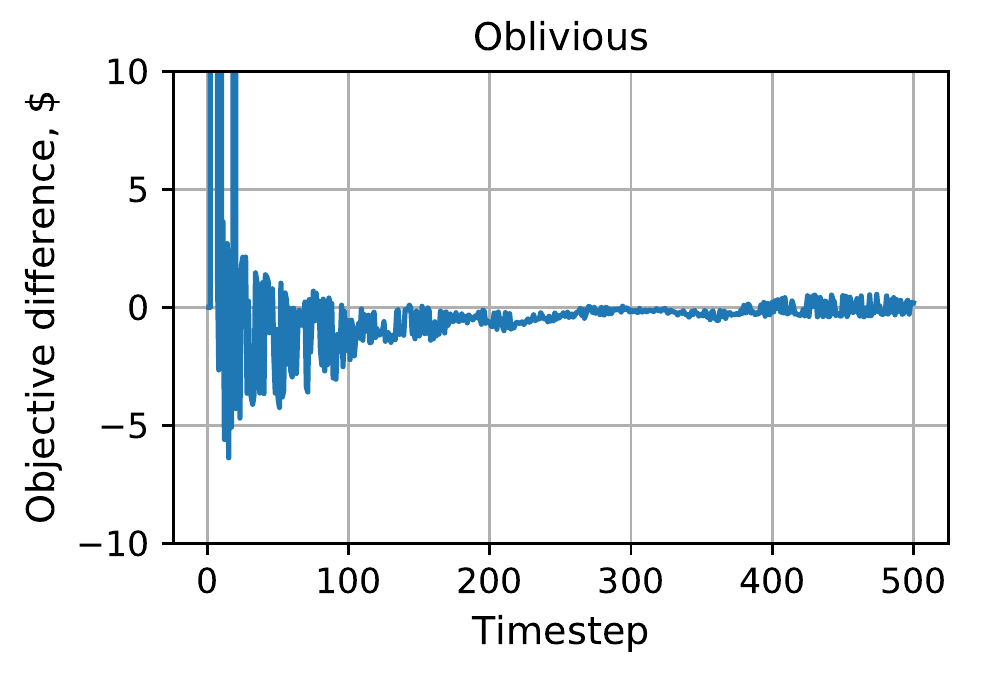}
\caption{Difference  in  the  DSO objective function between  the oblivious and oracular cases  for the 141-node test system.}
\label{fig:oracle_diff_141}
\end{figure}

{\color{black} 
\subsection{Scalability and Computational Performance}

To evaluate scalability  to larger distribution systems, we use the  141-node test system from  \cite{khodr2008maximum} and additionally populate it with controllable generators at nodes 30, 40, 50, 60, 70, 80, 101 and 121 with the production cost  in the range $\$10-\$17$ per MWh.
In the following, we use the fully constrained DRCC-OPF  since it is the most computationally demanding case.  All simulations were carried out on a PC with an Intel Core i7 processor with 2.50 GHz and 8 GB of memory.
Table \ref{tab:comp_times} compares the   computing times for 15- and 141-node test systems. In our case study we did not observe any computational abnormalities. 

Fig.~\ref{fig:oracle_diff_141} shows the difference between the objective functions in the oracle  and oblivious cases. As the number of time steps increases, so does the difference between the objective functions. As compared to the results in Fig.~\ref{fig:oracle_diff}(b) for the 15-node system, the convergence of the proposed learning scheme is similar in relative terms, but the residual differences are greater for the same time intervals due to a higher value of the objective function. 
The median DR utilization factor is \unit[58.26]{\%} of the available DR capacity in the system. We also observe that some fairly cheap DR flexibility and controllable generators  are blocked by the voltage and flow limits. The regret performance is similar to the 15-node test system showing a logarithmic progression. 
For instance, the average observed regret per time step is $\zeta^{[\text{obs}]} = \unit[22.14]{\$^2}$ in the first 10 time steps and it reduces to $\zeta^{[\text{obs}]} =  \unit[1.03]{\$^2}$ for subsequent time steps (11 to 500).  
}

\begin{table}[]
\renewcommand{\arraystretch}{1.3} 
\caption{Computing Times, s}
\label{tab:comp_times}
    \centering
    {\color{black}
    \begin{tabular}{m{3.3cm}|cc}
        \toprule
        & 15-node system & 141-node system \\
        \midrule
        Average computing time per time step & \unit[0.019]{} & \unit[2.427]{} \\
        Standard deviation of the computing time per time step&  \unit[0.006]{}  & \unit[0.262]{} \\
        \bottomrule
    \end{tabular}
    }
\end{table}

\vspace{-3mm}
\section{Conclusion}

This paper describes a learning approach that is capable of learning price sensitivities of residential DR resources and improves the utilization of these resources in the distribution system. The approach connects the least-square estimator and distributionally robust chance-constrained optimal power flow model that co-optimizes DR resources on a par with other dispatchable resources, while respecting operating limits on the distribution system. As the learning approach progresses, it reduces the systematic error inflicted by insufficient knowledge about price sensitivities of DR participants from the DSO perspective. The case study describes the usefulness of the proposed learning approach for different network instances.

\vspace{-3mm}

\bibliographystyle{IEEEtran}
\bibliography{literature_formated.bib,literature_local}

\newpage
\appendix
\numberwithin{equation}{subsection}
\subsection{Least Square Estimator (LSE)}
\label{ax:lse_derivation}

{\color{black} 
The LSE for computing $\hat{\beta}_{1,i}^{(t)}$ and $\hat{\beta}_{0,i}^{(t)}$ in  \eqref{eq:lse_beta_1} and \eqref{eq:lse_beta_2} is derived by  minimizing the sum of the squared errors:
\begin{align}
    \min_{\hat{\beta}_{1,i}^{(t)}, \hat{\beta}_{0,i}^{(t)}} \sum_{\tau=1}^{t-1} \left(2\hat{\beta}_{1,i}^{(t)} \lambda_{i,\tau} + \hat{\beta}_{0,i}^{(t)} - x_{i,\tau}\right)^2.
\end{align}
The resulting first-order optimality conditions are given by:
\begin{align}
    \left(\hat{\beta}_{1,i}^{(t)}\right): \quad &\sum_{\tau=1}^{t-1} \left( 2\hat{\beta}_{1,i}^{(t)} \lambda_{i,\tau}^2 + \hat{\beta}_{0,i}^{(t)} \lambda_{i,\tau} - \lambda_{i,\tau} x_{i,\tau}\right) = 0, \label{eq:lse_derivation_beta1_foc}\\
    \left(\hat{\beta}_{0,i}^{(t)}\right): \quad & \sum_{\tau=1}^{t-1} \left( \hat{\beta}_{0,i}^{(t)} + 2\hat{\beta}_{1,i}^{(t)} \lambda_{i,\tau} - x_{i,\tau}\right) = 0. \label{eq:lse_derivation_beta0_foc}
\end{align}
By using $\sum_{\tau=1}^{t-1} x_{i,\tau} = (t-1)\bar{x}_{i,t}$ and $\sum_{\tau=1}^{t-1} \lambda_{i,\tau} = (t-1)\bar{\lambda}_{i,t}$, we can insert \eqref{eq:lse_derivation_beta0_foc} into \eqref{eq:lse_derivation_beta1_foc} to obtain:
\begin{align}
    & \hat{\beta}_{1,i}^{(t)} = \frac{\sum_{\tau=1}^{t-1}\lambda_{i,\tau} x_{i,\tau} - (t-1)\bar{\lambda}_{i,t}\bar{x}_{i,t}}{2\sum_{\tau=1}^{t-1}\lambda_{i,\tau} \lambda_{i,\tau}^2 - (t-1)\bar{\lambda}_{i,t}^2} \\
    & \hat{\beta}_{0,i}^{(t)} = \bar{x}_{i,t} - \hat{\beta}_{1,i}^{(t)} \bar{\lambda}_{i,t}.
\end{align}
By recasting $\sum_{\tau=1}^{t-1}\lambda_{i,\tau} x_{i,\tau} - (t-1)\bar{\lambda}_{i,t}\bar{x}_{i,t}$ into $\sum_{\tau=1}^{t-1}(\lambda_{i,\tau} - \bar{\lambda}_{i,t})(x_{i,\tau} - \bar{x}_{i,t})$ and $\sum_{\tau=1}^{t-1}\lambda_{i,\tau} \lambda_{i,\tau}^2 - (t-1)\bar{\lambda}_{i,t}^2$ into $\sum_{\tau=1}^{t-1}(\lambda_{i,\tau} - \bar{\lambda}_{i,t})^2$ we obtain  the expressions \eqref{eq:lse_beta_1} and \eqref{eq:lse_beta_2}.
}

\subsection{Distributionally Robust Chance Constraints}
\label{ax:comlete_set_of_dr_chance_constraints}

Following the procedure described in Section~\ref{sec:dro}, semidefinite equivalents are obtained for:

\subsubsection{Equations \eqref{eq:fist_cc} and \eqref{eq:cc_gQ_up}, $\forall t \in \cal{T}, \forall \text{\textit{i}} \in \cal{N}^+$:}
\begin{subequations}
\begin{align}
     & M^{g\cdot,max}_{i,t} \succeq 0, \qquad \label{eq:app1}
     s^{g\cdot,max}_{i,t} + \frac{1}{\eta_g}\scal{\hat{\Omega}^{(t)}, M^{g\cdot,max}_{i,t}} \leq 0,  \\
     &M^{g,max}_{i,t} - \begin{bmatrix}
             0 & -\frac{1}{2} \alpha_i e \\
             -\frac{1}{2} \alpha_i e^\top &  g^\cdot_{i,t} - g_i^{\cdot,max} - s^{g\cdot,max}_{i,t}
         \end{bmatrix} 
         \succeq 0,
\end{align}
\end{subequations}
where ($\cdot$) is a placeholder for $P/Q$ denoting active and reactive power respectively.

\subsubsection{Equations \eqref{eq:cc_gP_low} and \eqref{eq:cc_gQ_low}, $\forall t \in \cal{T}, \forall \text{\textit{i}} \in \cal{N}^+$:}
\begin{subequations}
\begin{align}
     & M^{g\cdot,min}_{i,t} \succeq 0, \qquad s^{g\cdot,min}_{i,t} + \frac{1}{\eta_g}\scal{\hat{\Omega}^{(t)}, M^{g\cdot,min}_{i,t}} \leq 0, \\
     &M^{g\cdot,min}_{i,t} - \begin{bmatrix}
             0 & -\frac{1}{2} \alpha_i e \\
             -\frac{1}{2} \alpha_i e^\top &  -g^\cdot_{i,t} + g_i^{\cdot,min} - s^{g\cdot,min}_{i,t}
         \end{bmatrix} 
         \succeq 0,
\end{align}
\end{subequations}
where ($\cdot$) is a placeholder for $P/Q$ denoting active and reactive power respectively.

\subsubsection{Equation \eqref{eq:lower_voltage}, $\forall t \in \cal{T}, \forall \text{\textit{i}} \in \cal{N}^+$:}
\begin{subequations}
\begin{align}
     & M^{v,min}_{i,t} \succeq 0, \qquad s^{v,min}_{i,t} + \frac{1}{\eta_v}\scal{\hat{\Omega}^{(t)}, M^{v,min}_{i,t}} \leq 0, \\
     &M^{v,min}_{i,t} - \begin{bmatrix}
             0 & \frac{1}{2}T_i(\alpha)^\top \\
             \frac{1}{2}T_i(\alpha) &  -u_{i,t} + u_i^{min} - s^{v,min}
         \end{bmatrix} 
         \succeq 0,  \label{eq:append}
\end{align}
\end{subequations}

Eq.~\eqref{eq:app1}-\eqref{eq:append} are semidefinite constraints that can be solved efficiently using existing of-the-shelf solvers as explained in \cite{Esfahani2015Data}.

{\color{black} 
\subsection{Regret Components}
\label{ax:regret_components}
Using \eqref{eq:expected_costs}, \eqref{exp:sale_cost} and \eqref{eq:exp_dr_cost}, the expected cost as given by \eqref{eq:expected_cost} are reformulated as:
\begin{equation}
\begin{split}
    \Eptn[\bm{C}_t] =& \overbrace{\sum_{i\in\set{G}} (c_i(g_{i,t}^P)}^{\text{Exp. Generation Cost}} +  \overbrace{c_{i2}\alpha_i^2e^{\!\top}\hat{\Sigma}^{(t)}e)}^{\text{Exp. Balancing Cost}} \\
    & + \underbrace{\sum_{i\in\set{N}} \frac{1}{2\hat{\beta}_{1,i}} (x_{i,t}^*)^2 - \frac{\hat{\beta}_{1,i}}{2\hat{\beta}_{1,i}} x_{i,t}^* + \kappa_t \sum_{i\in\set{N}} x_{i,t}^*}_{\text{Exp. total Cost of DR}}
\end{split}
\end{equation}
Since there is no parameter uncertainty in the cost of generation, expected regret due to the expected energy provision is computed as:
\begin{align}
\zeta^{\text{[en]}}(t) &\coloneqq \sum_{i\in\set{N}} \left(\frac{1}{2\hat{\beta}_{1i}^{(t)}}(x_{i,t}^*)^2 - (\frac{\hat{\beta}_{0,i}^{(t)}}{2\hat{\beta}_{1,i}^{(t)}} - \kappa_t)x_{i,t}^*\right)  \nonumber \\ 
    & \quad - \sum_{i\in\set{N}} \left(\frac{1}{2\beta_{1i}}(x_{i,t}^*)^2 - (\frac{\beta_{0,i}}{2\beta_{1,i}} - \kappa_t)x_{i,t}^*\right) \\
    & = \sum_{i\in\set{N}} \left((\frac{1}{2\hat{\beta}_{1,i}^{(t)}} - \frac{1}{2\beta_{1,i}})(x_{i,t}^*)^2 - (\frac{\hat{\beta}_{0,i}^{(t)}}{2\hat{\beta}_{1,i}^{(t)}} - \frac{\beta_{0,i}}{2\beta_{1,i}}) x_{i,t}^*\right).\nonumber
\end{align}%
Similarly, the expected regret due to the expected cost of balancing is computed as:
\begin{equation}
\begin{split}
   \zeta^{\text{[bal]}}(t) &\coloneqq \sum_{i\in\set{N}} \left( \alpha_i^2 c_{2i} (e^{\!\top}\hat{\Sigma}^{(t)}e \right) - \sum_{i\in\set{N}} \left( \alpha_i^2 c_{2i} e^{\!\top}\Sigma e\right) \\
  &= \sum_{i\in\set{N}} 
\left( \alpha_i^2 c_{2i} e^{\!\top}(\hat{\Sigma}^{(t)} - \Sigma )e \right).
\end{split}%
\end{equation}%
Thus, the total regret at every time step is computed as:
\begin{equation}
    \zeta(t)=\zeta^{\text{[en]}}(t) + \zeta^{\text{[bal]}}(t).
\end{equation}
}

\end{document}